\documentclass{emulateapj}
\usepackage{apjfonts}
\usepackage{epsf}

\shorttitle{GALAXY FORMATION WITH DYNAMICAL RESPONSE}
\shortauthors{NAGASHIMA \& YOSHII}

\begin{document}

\title{Hierarchical Formation of Galaxies with Dynamical Response to 
Supernova-Induced Gas removal
}

\author{Masahiro Nagashima
\altaffilmark{1}
}
\affil{Division of Theoretical Astrophysics, National Astronomical
Observatory, Mitaka, Tokyo 181-8588, Japan\\
Department of Physics, University of Durham, South Road, Durham DH1 3LE, U.K.
}
\email{masa@scphys.kyoto-u.ac.jp}
\and
\author{Yuzuru Yoshii}
\affil{
Institute of Astronomy, School of Science, The University of
Tokyo, Mitaka, Tokyo 181-0015, Japan\\
Research Center for the Early Universe, School of Science, The
University of Tokyo, Bunkyo-ku, Tokyo 113-0033, Japan}

\altaffiltext{1}{Current address: Department of Physics, Graduate School of
Science, Kyoto University, Sakyo-ku, Kyoto 606-8502, Japan}

\begin{abstract}
 We reanalyze the formation and evolution of galaxies in the
 hierarchical clustering scenario.  Using a semi-analytic model (SAM) of
 galaxy formation described in this paper, which we hereafter call the
 Mitaka model, we extensively investigate the observed scaling relations
 of galaxies among photometric, kinematic, structural and chemical
 characteristics.  In such a scenario, spheroidal galaxies are assumed
 to be formed by major merger and subsequent starburst, in contrast to
 the traditional scenario of monolithic cloud collapse.  As a new
 ingredient of SAMs, we introduce the effects of dynamical response to
 supernova-induced gas removal on size and velocity dispersion, which
 play an important role on dwarf galaxy formation.  In previous
 theoretical studies of dwarf galaxies based on the monolithic cloud
 collapse given by Yoshii \& Arimoto and Dekel \& Silk, the dynamical
 response was treated in the extremes of a purely baryonic cloud and a
 baryonic cloud fully supported by surrounding dark matter.  To improve
 this simple treatment, in our previous paper, we formulated the
 dynamical response in more realistic, intermediate situations between
 the above extremes.  While the effects of dynamical response depend on
 the mass fraction of removed gas from a galaxy, how much amount of the
 gas remains just after major merger depends on the star formation
 history.  A variety of star formation histories are generated through
 the Monte Carlo realization of merging histories of dark halos, and it
 is found that our SAM naturally makes a wide variety of dwarf galaxies
 and their dispersed characteristics as observed.  It is also found that
 our result strongly depends on the adopted redshift dependence of star
 formation timescale, because it determines the gas fraction in
 high-redshift galaxies for which major mergers frequently occur.  We
 test four star formation models. The first model has a constant
 timescale of star formation independent of redshift.  The last model
 has a timescale proportional to the dynamical timescale of the galactic
 disk.  The other models have timescales intermediates of these two.
 The last model fails to reproduce observations, because it predicts
 only little amount of the leftover gas at major mergers, therefore
 giving too weak dynamical response on size and velocity dispersion of
 dwarf spheroidals.  The models, having a constant timescale of star
 formation or a timescale very weakly dependent on redshift, associated
 with our SAM, succeed to reproduce most observations from giant to
 dwarf galaxies, except that the adopted strong supernova feedback in
 this paper does not fully explain the color--magnitude relation under
 the cluster environment and the Tully-Fisher relation.  A direction of
 overcoming this remaining problem is also discussed.
\end{abstract}

\keywords{cosmology: theory -- galaxies: dwarf -- galaxies: evolution --
  galaxies: formation -- large-scale structure of universe }

\section{Introduction}
Dwarf galaxies give us many useful insights on the galaxy formation.
They occupy a dominant fraction among galaxies in number and have a 
wide variety of structures and formation histories.  Furthermore it 
would be easier to understand how physical processes such as heating 
by supernova explosions affect their evolution, because of shallowness 
of their gravitational potential well, compared to massive galaxies 
that have more complex formation histories and physical processes.  
Equally important is that massive galaxies have evolved via continuous 
mergers and accretion of less massive galaxies, according to the recent 
standard scenario of large-scale structure formation in the universe 
with the cold dark matter (CDM). This indicates that it is essential 
to understand the formation of dwarf galaxies even in understanding 
the formation of massive galaxies, because they would be formed by 
mergers of their {\it building blocks} like dwarf galaxies.

Before the CDM model gained its popularity, traditional models of
monolithic cloud collapse such as the galactic wind model for elliptical
galaxies \citep{els62,l69,i77,s79,ys79,ay86,ay87,ka97} and the infall
model for spiral galaxies \citep{ayt91} had been widely used in analyses
of galaxy evolution.  These are simple but strong tools to explain
various observations and then have contributed to construction of a
basic picture of galaxy evolution.  Based on such traditional models,
\citet{ds86} systematically investigated many aspects of dwarf galaxies.
Especially they focused on the role of supernova (SN) feedback, that is,
heating up and sweeping out of the galactic gas by multiple SN
explosions, in the processes of formation and evolution of dwarf
galaxies.  Since a large amount of heated gas is expelled by the
galactic wind, the self-gravitating system expands as a result of
dynamical response to the gas removal \citep{h80, m83, v86}.  They
considered two limiting cases for the gas removal.  One is a purely
self-gravitating gas cloud, and another is a gas cloud embedded in a
dominant dark halo.  They showed that many observed scaling relations
are well explained by their model.  Combined with the evolutionary
population synthesis code, \citet{ya87} extended their analysis to
estimating directly observable photometric properties, while only
considering the purely self-gravitating gas cloud.

Since the hierarchical clustering of dark halos predicted by the CDM
model becomes a standard structure formation scenario, the galaxy
formation scenario must be modified so as to be consistent with the
hierarchical merging of dark halos.  Explicitly taking into account the
Monte Carlo realization of merging histories of dark halos based on the
distribution function of initial density fluctuations, the so-called
semi-analytic models (SAMs) of galaxy formation have been developed
\citep[e.g.,][]{kwg93, c94, bcf96, ngs99, sp99, spf01, ntgy01, nytg02}.
SAMs include several important physical processes such as star
formation, supernova feedback, galaxy merger, population synthesis and
so on.  While SAMs well reproduce many observed properties of galaxies,
most of analyses have been limited to those of massive galaxies.  Thus
we focus on the formation of dwarf galaxies in the framework of the SAMs
in this paper.  In the analyses of dwarf galaxies, the dynamical
response to gas removal, which is known to play an important role as
shown by \citet{ds86} and \citet{ya87}, must be taken into account.
While self-consistent equilibrium models taking into account baryon and
dark matter have been analyzed by \citet{ys87} and \citet{cp92}, we
focus on the dynamical response of baryons within a dark matter halo.
Since we have derived the mathematical formula of dynamical response for
the galaxies consisting of baryon and dark matter in \citet{ny03}, it is
possible to incorporate them into our SAM.

The main purpose of this paper is a reanalysis of \citet{ds86} and
\citet{ya87} in the framework of SAMs.  Thus we mainly focus on the
formation of elliptical galaxies, which are assumed to be formed by
major merger and subsequent starburst.  In order to do this, we
construct the Mitaka model, which is a SAM including the effects of
dynamical response.  So far many scaling relations among photometric,
structural, and kinematical parameters of elliptical galaxies have been
observed, such as the color-magnitude relation \citep[e.g.,][]{b59}, the
velocity dispersion-magnitude relation \citep{fj76}, and the surface
brightness-size relation \citep{k77, kow83}.  To understand how these
relations are originated, especially on scales of dwarf galaxies, helps
to clarify the processes of galaxy formation and evolution in the
context of the cosmological structure formation scenario.  
Although some authors found by principal component analysis that
elliptical galaxies are distributed over the so-called fundamental 
plane in the three-dimensional space among photometric, structural and 
kinematical parameters \citep{wko85,d87,dd87}, we focus on the direct 
observables in this paper rather than their principal component 
projection.

This paper is outlined as follows.  In \S2 we describe our SAM.  In \S3
we constrain model parameters in our SAM using local observations.  In
\S4 we compare the theoretical predictions of SAM galaxies with various
observations.  In \S5 we examine consistency check of our SAM with other
observations.  In \S6 we show the cosmic star formation history.  In \S7
we discuss the nature of galaxies on the cooling diagram that has been
traditionally used for understanding the formation of galaxies.  In \S8
we provide summary and conclusion.

\section{Model}
The galaxy formation scenario that we use is as follows.  In the CDM
universe, dark matter halos cluster gravitationally and merge in a
manner that depends on the adopted power spectrum of the initial density
fluctuations.  In each of the merged dark halos, radiative gas cooling,
star formation, and gas reheating by supernovae occur.  The cooled dense
gas and stars constitute {\it galaxies}.  These galaxies sometimes merge
together in a common dark halo, and then more massive galaxies form.
Repeating these processes, galaxies form and evolve to the present
epoch.

Some ingredients of our SAM are revised.  Modifications include the
shape of mass function of dark halos, the star formation (SF) timescale,
the merger timescale of galaxies taking into account the tidal stripping
of subhalos, and the dynamical response to gas removal caused by
starburst during major merger.  The details are described below.

\subsection{Merging Histories of Dark Halos}\label{sec:mh}
The merging histories of dark halos are realized by a Monte Carlo method
proposed by \citet{sk99}, based on the extended Press-Schechter (PS)
formalism \citep{bcek91, b91, lc93}.  This formalism is an extension of
the Press-Schechter formalism \citep{ps74}, which gives the mass
function of dark halos, $n(M)$, to estimate the mass function of
progenitor halos with mass $M_{1}$ at a redshift $z_{0}+\Delta z$ of a
single dark halo with mass $M_{0}$ collapsing at a redshift $z_{0}$,
$n(M_{1}; z_{0}+\Delta z|M_{0}; z_{0})dM_{1}$.  According to this mass
function, a set of progenitors is realized.  By repeating this, we
obtain a {\it merger tree}.  Realized trees are summed with a weight
given by a mass function at output redshift.  Dark halos with circular
velocity $V_{\rm circ}\geq V_{\rm low}=30$km~s$^{-1}$ are regarded as
isolated halos, otherwise as diffuse accreted matter.

\begin{figure}
\plotone{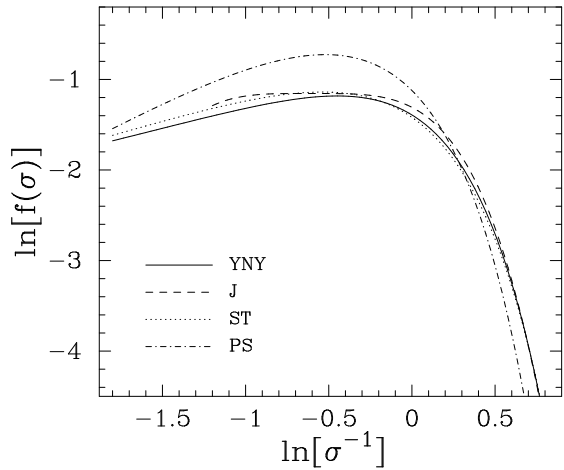}

\caption{Mass functions of dark halos.  The solid line represents the
 mass function given by \citet[][YNY]{yny03}, which is used in this paper.  The
 dot-dashed line is the original PS mass functions, respectively.  The
 dotted line is the analytic result by \citet[][ST]{st99}.  The dashed line is
 the $N$-body result over the resolved range by \citet[][J]{j01}.  }

\label{fig:massfn}
\end{figure}

In our previous papers, we adopted the PS mass function to provide the
weight for summing merger trees.  Recent high-resolution $N$-body
simulations, however, suggest that the PS mass function should be
slightly corrected \citep[e.g.,][]{j01}.  According to the notation in
\citet{j01}, the original PS mass function is written by
\begin{equation}
 f(\sigma;{\rm PS})=\sqrt{\frac{2}{\pi}}\frac{\delta_{c}}{\sigma}
\exp\left(-\frac{\delta_{c}^{2}}{2\sigma^{2}}\right),
\end{equation}
and the cumulative mass function $n(M)$ is related with the above
function by
\begin{equation}
 f(\sigma;X)=\frac{M}{\rho_{0}}\frac{dn(M)}{d\ln\sigma^{-1}},
\end{equation}
where $X$ specifies a model such as PS, $\rho_{0}$ is the mean density
of the universe, $\sigma$ denotes the standard deviation of the density
fluctuation field, and $\delta_{c}$ is the critical density contrast for
collapse assuming spherically symmetric collapse \citep{t69, gg72}.  In
this paper we use the following mass function given by
\citet[][hereafter YNY]{yny03}
instead of the PS mass function,
\begin{equation}
 f(\sigma;{\rm YNY})=A(1+x^{C})x^{D}\exp(-x^{2}),
\end{equation}
where $x=B\delta_{c}/\sqrt{2}\sigma$,
$A=2/[\Gamma(D/2)+\Gamma(\{C+D\}/2)], B=0.893, C=1.39$ and $D=0.408$.
This is a fitting function that satisfies the normalization condition,
that is, the integration over all rage of $\nu$ is unity.  These
functions are plotted in Figure \ref{fig:massfn} by the solid line (YNY)
and the dot-dashed line (PS).  We also show other often used formula
given by \citet{st99}, 
\begin{equation}
 f(\sigma;{\rm ST})=A\sqrt{\frac{2a}{\pi}}\left[1+\left(\frac{\sigma^{2}}{a\delta_{c}^{2}}\right)^{p}\right]\frac{\delta_{c}}
{\sigma}\exp\left(-\frac{a\delta_{c}^{2}}{2\sigma^{2}}\right),
\end{equation}
and by \citet{j01},
\begin{equation}
 f(\sigma;{\rm J})=0.315\exp(-|\ln\sigma^{-1}+0.61|^{3.8}),
\end{equation}
where $A=0.3222, a=0.707$ and $p=0.3$ for the former (ST)
and the latter (J) formula is valid over the range
$-1.2\leq\ln\sigma^{-1}\leq 1.05$.  

The number density of dark halos with large mass given by the above
three functions (YNY, ST and J) and by the $N$-body simulation are
similar to those given by analytic estimation \citep[e.g.,][]{yng96,
monaco98, n01} and are therefore trustworthy.  With $N$-body
simulations, it is difficult to estimate the mass function at low mass
because of the limited resolution and the uncertainty in the
identification of dark halos.  The very high resolution $N$-body
simulations, made using the adaptive mesh refinement method by
\citet{yy01} and \citet{y02}, predict a mass function similar to that of
ST.  This mass function extends to a mass 10 times smaller than that of
\citet{j01} and has a slightly different slope.  Thus we adopt the YNY
mass function as above.  A comparison between the YNY mass function and
the $N$-body simulation is discussed in a separate paper \citep{yny03}.
In \S\ref{sec:paramset} we show how the mass function of dark halos
affects the luminosity function of galaxies.

In short, we realize merger trees by using the extended PS formalism for
dark halos whose mass function is given by the YNY at each output
redshift.  

In this paper, we only consider a recent standard $\Lambda$CDM model,
that is, $\Omega_{0}=0.3, \Omega_{\Lambda}=0.7, h=0.7$ and
$\sigma_{8}=0.9$, where these parameters denote the mean density of the
universe, the cosmological constant, the Hubble parameter and the
normalization of the power spectrum of the initial density fluctuation
field.  The shape of the power spectrum given by \citet{s95}, which is a
modified one of \citet[][hereafter BBKS]{bbks} taking into account the
baryonic effects, is adopted.

\subsection{Tidal Stripping of Subhalos}\label{sec:tidal}
Most of recent high resolution $N$-body simulations suggest that
swallowed dark halos survive in their host halo as {\it subhalos}.
Envelopes of those subhalos are stripped by tidal force from the host
halo.  We assume the radius of a tidally stripped subhalo $r_{\rm t}$
by
\begin{equation}
 \frac{r_{\rm t}}{r_{\rm s}}=\frac{r_{\rm peri}}{r_{\rm apo}}\frac{M_{\rm h}}{M_{\rm s}}\left(\frac{V_{\rm circ,s}}{V_{\rm 
circ,h}}\right)^{3},
\end{equation}
where $r_{\rm peri}$ and $r_{\rm apo}$ are the pericenter and apocenter
for the orbit of the subhalo, respectively, and subscripts ``h'' and
``s'' indicate the host halo and subhalo, respectively.  In this paper,
a ratio of $r_{\rm peri}/r_{\rm apo}=0.2$ is assumed \citep{gmglqs98,
oh99, oh00}.  Because a singular isothermal profile for subhalos is
assumed, their mass decreases proportional to $r_{\rm t}/r_{\rm s}$.
The mass of stripped subhalos is used in the estimation of dynamical
response to gas removal on the size and velocity dispersion when
satellite galaxies merge together (see \S\S\ref{sec:response}).

\subsection{Gas Cooling, Star Formation and Supernova Feedback}
The mean mass density in dark halos is assumed to be proportional to the
cosmic mean density at the epoch of collapse using a spherically
symmetric collapse model \citep{t69, gg72}.  Each collapsing dark halo
contains baryonic matter with a mass fraction $\Omega_{\rm
b}/\Omega_{0}$, where $\Omega_{\rm b}$ is the baryon density parameter.
We adopt a value of $\Omega_{\rm b}h^{2}=0.02$ that is recently
suggested by the BOOMERANG Project measuring the anisotropy of the
cosmic microwave background \citep{boomerang}.  This value is an
intermediate one between $(0.64-1.4)\times 10^{-2}$ given by analysis of
light element abundance produced by big bang nucleosynthesis
\citep{syb00} and $(2.24\pm 0.09)\times 10^{-2}$ given by analysis of
cosmic microwave background observed by $WMAP$ \citep{s03}.  The
baryonic matter consists of diffuse hot gas, dense cold gas, and stars.

When a halo collapses, the hot gas is shock-heated to the virial
temperature of the halo with an isothermal density profile.  A part of
the hot gas cools and accretes to the disk of a galaxy until subsequent
collapse of dark halos containing this halo.  The amount of the cold gas
involved is calculated by using metallicity-dependent cooling functions
provided by \citet{sd93}.  The difference of cooling rates between the
primordial and metal-polluted gases is prominent at $T\sim 10^{6}$K due
to line-cooling of metals.  Chemical enrichment in hot gas is
consistently solved with star formation and SN feedback.  The cooling
is, however, very efficient in dark halos with a virial temperature of
$T\sim 10^{6}$K even in the case of the primordial gas, so the
metallicity dependence of cooling rate only slightly affects our
results.  In order to avoid the formation of unphysically large
galaxies, the cooling process is applied only to halos with $V_{\rm
circ}\leq V_{\rm cut}=$250 km~s$^{-1}$.  This manipulation would be
needed, because the simple isothermal distribution forms so-called
``monster galaxies'' due to too efficient cooling at the center of
halos.  While \citet{c00} adopted another isothermal distribution with
central core instead of such a simple cutoff of the cooling and
\citet{bbflbc03} considered some additional mechanisms such as the
heating of hot gas by SNe and by heat conduction from outside as well as
its removal by superwinds from halos, we take the above simple approach.
The value of $V_{\rm cut}$ is rather small compared with our previous
paper and other SAMs.  This is caused by our assumption that invisible
stars have negligible fraction, which is introduced to darken luminosity
of galaxies (\S\S\ref{sec:photo}).  This smaller value of $V_{\rm cut}$
makes the color of large galaxies less red, which shows up on a bright
portion of the color-magnitude relation of elliptical galaxies
(\S\S\ref{sec:cmr}).

Stars in disks are formed from the cold gas.  The SF rate (SFR)
$\dot{M}_{*}$ is given by the cold gas mass $M_{\rm cold}$ and a SF
timescale $\tau_{*}$ as $\dot{M}_{*}=M_{\rm cold}/\tau_{*}$.  Now we
consider two SF models.  One is a constant star formation (CSF), in
which $\tau_{*}$ is constant against redshift.  Another is a dynamical 
star formation (DSF), in which $\tau_{*}$ is proportional to the 
dynamical timescale of the halo, which allows for the possibility
that the SF efficiency is variable with redshift.  We then express these
SF timescales as
\begin{eqnarray}
\tau_{*}=\left\{
\begin{array}{ll}
\displaystyle{\tau_{*}^{0}[1+\beta(V_{\rm circ})]} & \mbox{(CSF)},\\
\displaystyle{\tau_{*}^{0}[1+\beta(V_{\rm circ})]
\left[\frac{\tau_{\rm dyn}(z)}{\tau_{\rm dyn}(0)}\right]
(1+z)^{\sigma}} & \mbox{(DSF)},
\end{array}
\right.
\label{eqn:sft}
\end{eqnarray}
where $\tau_{*}^{0}$ and $\sigma$ are free parameters, and $\beta$
indicates the ratio of the SF timescale to the reheating timescale by the
SN feedback defined by equation (\ref{eqn:beta}) (see below).  Pure DSF
occurs when $\sigma=0$.  Because the timescale of cold gas
consumption is equal to $\tau_{*}/(1+\beta-R)$, where $R$ is the
returned mass fraction from evolved stars ($R=0.25$ in this paper), the
mass fraction of cold gas in galaxies that is nearly constant against
their magnitude is automatically adjusted by multiplying $(1+\beta)$.
Hence the parameter $\alpha_{*}$ originally introduced by \citet{c94} is
eliminated by introducing the factor $(1+\beta)$.
The parameter $\tau_{*}^{0}$ is so chosen as to match the mass 
fraction of cold gas with the observed fraction (see \S 3).  
Thereby the SF-related parameters are constrained according to 
\citet{c00}.  Since not all of cold gas might be observed, the 
observed data give a lower limit to mass fraction of cold gas.  

\begin{figure}
\plotone{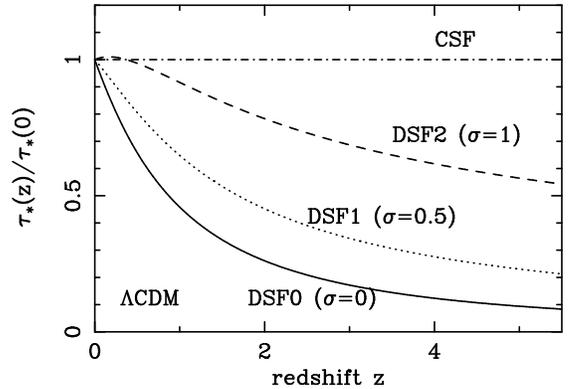}

\caption{Star formation timescales $\tau_{*}$ as a function of redshift
$z$ in a $\Lambda$CDM model.  The timescale is normalized to unity at
the present epoch.  The dot-dashed line represent the constant SF model
(CSF).  The solid, dotted and dashed lines represent the variants of
dynamical SF model with $\sigma=0$ (DSF0), 0.5 (DSF1) and 1 (DSF2),
respectively, according to the prescriptions in equation \ref{eqn:sft}.  } 
\label{fig:sfr}

\end{figure}

Figure \ref{fig:sfr} shows the redshift dependence of SF timescale 
for the four SF models of CSF (dot-dashed line), DSF0 ($\sigma=0$; 
solid line), DSF1 ($\sigma=0.5$; dotted line) and DSF2 ($\sigma=1$; 
dashed line).  Objects, which collapse at higher redshift, have higher 
density and therefore shorter dynamical timescale.  Evidently, the 
DSF with smaller $\sigma$ gives more rapid conversion of cold gas 
into stars, compared with the CSF.  The four SF models predict 
different mass fraction of cold gas at high redshift, leading to
quite different characteristics of dwarf galaxies.  This indicates, 
as will be clarified later, that the redshift dependence of SF 
timescale can be constrained particularly from observed structures 
and photometric properties of dwarf galaxies.

Massive stars explode as Type II SNe and heat up the surrounding cold
gas.  This SN feedback reheats the cold gas at a rate of $\dot{M}_{\rm
reheat}={M_{\rm cold}}/{\tau_{\rm reheat}}$, where the timescale of
reheating is given by
\begin{equation}
\tau_{\rm reheat}=\frac{ \tau_{*}}{\beta(V_{\rm circ})},
\end{equation}
where
\begin{equation}
\beta(V_{\rm circ})\equiv\left(\frac{V_{\rm circ}}{V_{\rm hot}}
\right)^{\alpha_{\rm hot}}.
\label{eqn:beta}
\end{equation}
The free parameters $V_{\rm hot}$ and $\alpha_{\rm hot}$ are determined
by matching the local luminosity function of galaxies with observations.

With the above equations and parameters, we obtain the masses of hot
gas, cold gas, and disk stars as a function of time or redshift.
Chemical enrichment is also taken into account, adopting the {\it
heavy-element yield} of $y=2Z_{\odot}$, assuming the instantaneous
recycling approximation with the returned mass fraction from evolved
stars $R=0.25$.  All of newly produced metals are released into cold
gas, then by SN feedback, a part of them is expelled into hot gas.  Some
metals in hot gas are brought back to cold gas by subsequent cooling,
and are accumulated in stars by their formation.

\subsection{Mergers of dark halos and galaxies}
When two or more progenitor halos have merged, the newly formed larger
halo should contain at least two or more galaxies which had originally
resided in the individual progenitor halos.  By definition, we identify
the central galaxy in the new common halo with the central galaxy
contained in the most massive one of the progenitor halos.  Other
galaxies are regarded as satellite galaxies.  These satellites merge by
either dynamical friction or random collision.  The timescale of merging
by dynamical friction is given by $\tau_{\rm mrg}=f_{\rm mrg}\tau_{\rm
fric}$, where $\tau_{\rm fric}$ is given by \citet{bt87}, which is
estimated from masses of the new common halo and the tidally truncated
subhalo.  The parameter $f_{\rm mrg}$ is set to 0.7 in this paper.  When
the time elapsed after merging of a progenitor halo exceeds $\tau_{\rm
mrg}$, the satellite galaxy is accreted to the central galaxy.  On the
other hand, the mean free timescale of random collision of satellite
galaxies $\tau_{\rm coll}$ is given by \citet{mh97}.  With a probability
$\Delta t/\tau_{\rm coll}$, where $\Delta t$ is the time step
corresponding to the redshift interval $\Delta z$ of merger tree of dark
halos, a satellite galaxy merges with another satellite picked out
randomly \citep{sp99}.

Consider the case when two galaxies of masses $m_1$ and $m_2 (>m_1)$
merge together.  If the mass ratio $f=m_1/m_2$ is larger than a certain
critical value of $f_{\rm bulge}$, we assume that a starburst occurs and
that all of the cold gas turns into stars and hot gas, which fills the
resulting halo, and all of the stars populate the bulge of a new galaxy.
On the other hand, if $f<f_{\rm bulge}$, no starburst occurs, and a
smaller galaxy is simply absorbed into the disk of a larger galaxy.
Throughout this paper we use $f_{\rm bulge}=0.5$, which gives a
consistent morphological fraction in galaxy number counts.

\subsection{Size of Galaxies and Dynamical Response to Starburst-induced Gas Removal}\label{sec:response}

We assume that the size of spiral galaxies is determined by a radius at
which the gas is supported by rotation, under the conservation of
specific angular momentum of hot gas that cools and contracts.  We also
assume that the initial specific angular momentum of the gas is the same
as that of the host dark halo.  Acquisition of the angular momentum of
dark halos is determined by tidal torques in the initial density
fluctuation field \citep{w84, ct96a, ct96b, ng98}.  The distribution of
the dimensionless spin parameter $\lambda_{H}$, which is defined by
$\lambda_{H}\equiv L|E|^{1/2}/GM^{5/2}$ where $L$ is the angular
momentum and $E$ is the binding energy, is well approximated by a
log-normal distribution \citep{mmw98},
\begin{equation}
 p(\lambda_{H})d\lambda_{H}=
\frac{1}{\sqrt{2\pi}\sigma_{\lambda}}
\exp\left[-\frac{(\ln\lambda_{H}-\ln\bar{\lambda})^2}
{2\sigma_{\lambda}^{2}}\right] d\ln\lambda_{H},
\label{eqn:spin}
\end{equation}
where $\bar{\lambda}$ is the mean value of spin parameter and
$\sigma_{\lambda}$ is its logarithmic variance.  We adopt
$\bar{\lambda}=0.03$ and $\sigma_{\lambda}=0.5$.  When the specific
angular momentum is conserved, the effective radius $r_{e}$ of a
presently observed galaxy at $z=0$ is related to the initial radius
$R_{i}$ of the progenitor gas sphere via
$r_{e}=(1.68/\sqrt{2})\lambda_{H}R_{i}$ \citep{f79, fe80, f83}.  The
initial radius $R_{i}$ is set to be the smaller one between the virial
radius of the host halo and the cooling radius.  A disk of a galaxy
grows due to cooling and accretion of hot gas from more distant envelope
of its host halo.  In our model, when the estimated radius by the above
equation becomes larger than that in the previous time-step, the radius
grows to the new larger value in the next step.  At that time, the disk
rotation velocity $V_{d}$ is set to be the circular velocity of its host 
dark halo.

Size estimation of high-redshift spiral galaxies, however, carries
uncertainties because of the large dispersion in their observed size
distribution.  For example, \citet{s99} suggests only mild evolution of
disk size against redshift, taking into account the selection effects 
arising from the detection threshold of surface brightness, although 
the above simple model predicts disk size proportional to virial radius 
$R_{\rm vir}$ of host dark halos evolving as $R_{\rm vir}\propto 1/(1+z)$ 
for fixed mass.  Allowing for the possibility that the conservation of 
angular momentum is not complete, we generalize this size estimation by 
introducing a simple redshift dependence,
\begin{equation}
 r_{e}=\frac{1.68}{\sqrt{2}}\lambda_{H}R_{i}(1+z)^{\rho},
\label{eqn:sizerho}
\end{equation}
where $\rho$ is a free parameter.  We simply use $\rho=1$ as a reference
value in this paper.  The effect of changing $\rho$ emerges in the
selection effects due to the cosmological dimming of surface brightness
and in the dust extinction, because the dust column density also changes
with galaxy size.  This is discussed in \S\ref{sec:consistency}.

Size of early-type galaxies, which likely form from galaxy mergers, are
primarily determined by the virial radius of the baryonic component.
When a major merger of galaxies occurs, assuming the energy
conservation, we estimate the velocity dispersion of the merged system.
Now we assign the subscript 0 to the merged galaxy, and subscripts 1 and
2 to the central and satellite galaxies, respectively, in the case of
central-satellite merger, or to larger and smaller galaxies,
respectively, in the case of satellite-satellite merger.  Using the
virial theorem, the total energy for each galaxy is
\begin{equation}
 E_{i}=-\frac{1}{2}[M_{b}V_{b}^{2}+(M_{d}+M_{\rm cold})V_{d}^{2}],
\end{equation}
where $M_{b}$ and $M_{d}$ are the masses of bulge and disk,
respectively, and $V_{b}$ and $V_{d}$ are the velocity dispersion of
bulge and the rotation velocity of disk, respectively.  Assuming the 
virial equilibrium, the binding energy $E_{b}$ between the progenitors 
just before the merger is given by
\begin{equation}
 E_{b}=-\frac{E_{1}E_{2}}{(M_{2}/M_{1})E_{1}+(M_{1}/M_{2})E_{2}}.
\end{equation}
Then we obtain
\begin{equation}
 E_{1}+E_{2}+E_{b}=E_{0}.
\end{equation}
Just after the merger there is only the bulge component consisting of
cold gas and stars in the merger remnant, whose velocity dispersion is
directly estimated from the above equation.  This procedure is similar
to \citet{c00}, although they argued it in terms of size estimation.  
Then the size of the system just after the merger is defined 
by
\begin{equation}
 r_{i}=\frac{GM_{i}}{2V_{b}^{2}},
\end{equation}
where $M_{i}=M_{*}+M_{\rm cold}$ is the total baryonic mass of the
merged system.  

Next, the cold gas turns into stars and hot gas.  Newly formed stellar
mass is nearly equal to $M_{\rm cold}/(1+\beta)$ and the rest of the
cold gas is expelled from the merged system to the halo by SN feedback.
The final mass after the mass loss, $M_{f}$, can be estimated from the
known $\beta$.  Assuming the density distributions of baryonic and dark 
matters, the dynamical response on the structural parameters to the mass 
loss can be estimated.  In this paper we adopt the Jaffe model 
\citep{jaffe} for baryonic matter and the
singular isothermal sphere for dark matter, and assume slow (adiabatic)
gas removal compared with dynamical timescale of the system.  Defining
the ratios of mass, size, density and velocity dispersion at final state
relative to those at initial state by ${\cal M}, R, Y$ and $U$, the 
response under the above assumption is approximately given by
\begin{eqnarray}
R&\equiv&\frac{r_{f}}{r_{i}}=\frac{1+D/2}{{\cal M}+D/2},\\
U&\equiv&\frac{V_{b,f}}{V_{b,i}}=\sqrt{\frac{YR^{2}+Df(z_{f})/2}{1+Df(z_{i})/2}},
\end{eqnarray}
where ${\cal M}=YR^{3}$, $D=1/y_{i}z_{i}^{2}$, $y$ and $z$ are the
ratios of density and size of baryonic matter to those of dark matter
and, $f(z)$ is a function defined in Appendix.  The subscripts $i$ and
$f$ stand for the initial and final states in the mass loss process.
The details are shown in Appendix and \citet{ny03}.  The parameter $D$
indicates the contribution of dark matter to the gravitational potential
felt by baryonic matter in the central region of a halo.  In actual
calculation, we use a circular velocity at the center of dark halos,
$V_{\rm cent}$ defined below, to estimate $D$ as $2V_{\rm
cent}^{2}/V_{b,i}^{2}$, instead of the ratios of size $y_{i}$ and
density $z_{i}$ which characterize the global property of dark halos.
If there is negligible dark matter ($D\to 0$), the well-known result of
adiabatic invariant $(M_{*}+M_{\rm cold})r$ emerges
\citep[e.g.,][]{ya87}.  In contrast, if there is negligible baryonic
matter ($D\to\infty$), $R$ and $U$ become unity, that is, size and
velocity dispersion do not change during mass loss.  The effect of the
dynamical response is the most prominent for dwarf galaxies of low
circular velocity.

Considering realistic situations, the baryonic matter often condenses in
the central region and becomes denser than the average density in the
halo.  It is likely that the depth of gravitational potential well is
changed when a part of baryonic mass is removed due to SN feedback.  To
take into account this process, we define a central circular velocity of
dark halo $V_{\rm cent}$.  When a dark halo collapses without any
progenitors, $V_{\rm cent}$ is set to $V_{\rm circ}$.  After that,
although the mass of the dark halo grows up by subsequent accretion
and/or mergers, $V_{\rm cent}$ remains constant or decreases by the
dynamical response.  When the mass is doubled, $V_{\rm cent}$ is set to
$V_{\rm circ}$ at that time again.  The dynamical response to mass loss
from a central galaxy of a dark halo by SN feedback lowers $V_{\rm
cent}$ of the dark halo as follows:
\begin{equation}
 \frac{V_{{\rm cent},f}}{V_{{\rm cent},i}}=
  \frac{M_{f}/2+M_{d}(r_{i}/r_{d})}{M_{i}/2+M_{d}(r_{i}/r_{d})}.
\end{equation}
The change of $V_{\rm cent}$ in each time step is only a few per cent.
Under these conditions the approximation of static gravitational
potential of dark matter is valid during starburst.

Once a dark halo falls into its host dark halo, it is treated as a
subhalo.  Because we assume that subhalos do not grow up in mass, the
central circular velocity of the subhalos monotonically decreases.  Thus
this affects the dynamical response later when mergers between satellite
galaxies occur.  We approximate that the resultant density distribution
remains to be isothermal with $V_{\rm cent}$ at least within the galaxy
size.

\subsection{Photometric Properties and Morphological Identification}\label{sec:photo}
The above processes are repeated until the output redshift and then the
SF history of each galaxy is obtained.  For the purpose of comparison
with observations, we use a stellar population synthesis approach, from
which the luminosities and colors of model galaxies are calculated.
Given the SFR as a function of time or redshift, the absolute luminosity
and colors of individual galaxies are calculated using a population
synthesis code by \citet{ka97}.  The stellar metallicity grids in the
code cover a range from $Z_{*}=$0.0001 to 0.05. Note that we now define
the metallicity as the mass fraction of metals.  The initial stellar mass
function (IMF) that we adopt is the power-law IMF of Salpeter form, with
lower and upper mass limits of $0.1M_{\odot}$ and $60M_{\odot}$,
respectively.

In most of SAM analyses, it has been assumed that there is a substantial
fraction of invisible stars such as brown dwarfs.  \citet{c94}
introduced a parameter defined as $\Upsilon=(M_{\rm lum}+M_{\rm
BD})/M_{\rm lum}$, where $M_{\rm lum}$ is the total mass of luminous
stars with mass larger than $0.1M_\odot$ and $M_{\rm BD}$ is that of
invisible brown dwarfs.  A range of $\Upsilon\sim 1-3$ has been assumed
depending on ingredients of SAMs.  For example, \citet{c00} assumed
$\Upsilon=3.07$ in the case of $\Omega_{\rm b}=0.04$.  In this paper,
however, we do not assume the existence of substantial fraction of
invisible stars.  If a large value of $\Upsilon$ is adopted, the
mass-to-light ratio of galaxies is too high to agree with observations.
Thus we fix $\Upsilon=1$.  \citet{sp99} also adopted a small value of
$\Upsilon=1.25$ in their $\Lambda$CDM.3 model for $\Omega_{\rm b}=0.037$
(in their notation $f^{*}_{\rm lum}=1/\Upsilon=0.8$).

The optical depth of internal dust is consistently estimated by our SAM.
We take the usual assumption that the abundance of dust is proportional
to the metallicity of cold gas, and then the optical depth is
proportional to the column density of metals.  Then the optical depth
$\tau$ is given by
\begin{equation}
 \tau\propto\frac{M_{\rm cold}Z_{\rm cold}}{r_{e}^{2}},
\label{eqn:dust}
\end{equation}
where $r_{e}$ is the effective radius of the galactic disk.  There are
large uncertainties in estimating the proportionality constant, but we
adopt about a factor of two smaller value compared with that in
\citet{c00}, otherwise it predicts too strong extinction to reproduce
galaxy number counts, presumably because our chemical yield is higher
than theirs.  Wavelength dependence of optical depth is assumed to be
the same as the Galactic extinction curve given by \citet{seaton79}.
Dust distribution is simply assumed to be the slab dust \citep{ddp89},
according to our previous papers.  We found that the resultant mean
extinction for spiral galaxies is close to a model by \citet{c00}.

We classify galaxies into different morphological types according to 
the $B$-band bulge-to-disk luminosity ratio $B/D$.  In this paper, 
following \citet{sdv86}, galaxies with $B/D\geq 1.52$, $0.68\leq 
B/D<1.52$, and $B/D<0.68$ are classified as elliptical, lenticular, 
and spiral galaxies, respectively.  \citet{kwg93} and \citet{bcf96} 
showed that this method of type classification well reproduces the 
observed type mix.

\section{Parameter Settings}\label{sec:paramset}
As already mentioned, we adopt a standard $\Lambda$CDM model.  The
cosmological parameters are $\Omega_{0}=0.3, \Omega_{\Lambda}=0.7,
h=0.7$ and $\sigma_{8}=0.9$.  The baryon density parameter $\Omega_{\rm
b}=0.02h^{-2}$ is used.

The astrophysical parameters are constrained from local observations,
according to the procedure discussed in \citet{ntgy01, nytg02}.  
The adopted values are slightly different from those in our previous 
papers.  This is mainly caused by adopting the different mass
function of dark halos (\S\S\ref{sec:mh}) and by fixing $\Upsilon=1$
(\S\S\ref{sec:photo}).  Values of these parameters are tabulated in
Tables \ref{tab:astro0} and \ref{tab:astro}.

\begin{deluxetable*}{llll}
\tabletypesize{\scriptsize}
\tablecaption{Standard Settings of Astrophysical Parameters in the
 Mitaka Model\label{tab:astro0}}  
\tablewidth{0pt}
\tablehead{
\colhead{Parameter} & \colhead{Value} & \colhead{Annotation} & \colhead{Observation}}
\startdata
\hspace{-2ex}
$\begin{array}{ll}
{V_{\rm hot}}\\
{\alpha_{\rm hot}}
\end{array}$
 & 
\hspace{-2ex}
$\left. \begin{array}{ll}
150 \mbox{km~s}^{-1} \\
4
\end{array}\right\}$
& supernova feedback-related (\S\S2.3) &
luminosity functions (Figure 3) \\
$V_{\rm cut}$ & 250 km~s$^{-1}$ & cooling cut-off (\S\S2.3) & luminosity
 functions (Figure 3)\\
$V_{\rm low}$ & 30 km~s$^{-1}$ & minimum circular velocity of dark halos
 (\S\S2.1) & ---\\
{$y$} & 2 $Z_{\odot}$ & heavy-element yield (\S\S2.3) & metallicity distribution
 (Figure 18)\\
{$f_{\rm bulge}$} & 0.5 & major/minor merger criterion (\S\S2.4) & morphological
counts\\
{$f_{\rm mrg}$} & 0.7 & coefficient of dynamical friction timescale
 (\S\S2.4) &
luminosity functions (Figure 3) \\
\hspace{-2ex}
$\begin{array}{ll}
{\bar{\lambda}}\\
{\sigma_{\lambda}}
\end{array}$
&
\hspace{-2ex}
$\left. \begin{array}{ll}
{0.03}\\
{0.5}
\end{array}\right\}$
&
spin parameter distribution (\S\S2.5) & disk size (Figure 5)\\
$\rho$ & 1 & redshift dependence of disk size (\S\S2.5) & faint galaxy
 number counts (Figures 20 and 21)\\
{$\Upsilon$} & 1 & fraction of invisible stellar mass (\S\S2.6) & mass-to-light
 ratio (Figure 17) \\
\enddata
\tablecomments{Cosmological parameters are: $(\Omega_{0},
\Omega_{\Lambda}, h, \sigma_{8}, \Omega_{\rm b})=(0.3, 0.7, 0.7, 0.9,
0.02h^{-2})$. } 
\end{deluxetable*}

\begin{deluxetable}{ccc}
\tabletypesize{\scriptsize}
\tablecaption{Star Formation-Model Dependent Parameters\label{tab:astro}}  
\tablewidth{0pt}
\tablehead{
\colhead{Model} & 
 \colhead{$\tau_{*}^{0}$ (Gyr)} & \colhead{$\sigma$}}
\startdata
CSF  & 1.3 & -- \\
DSF0 & 1.7 & 0  \\
DSF1 & 1.3 & 0.5\\
DSF2 & 1.5 & 1  \\
\enddata
\tablecomments{$\tau_{*}^{0}$ is determined by matching the mass fraction
 of cold gas with the observed fraction in spiral galaxies
 [equation \ref{eqn:sft} and Figure 4].}
\end{deluxetable}

\begin{figure}
\plotone{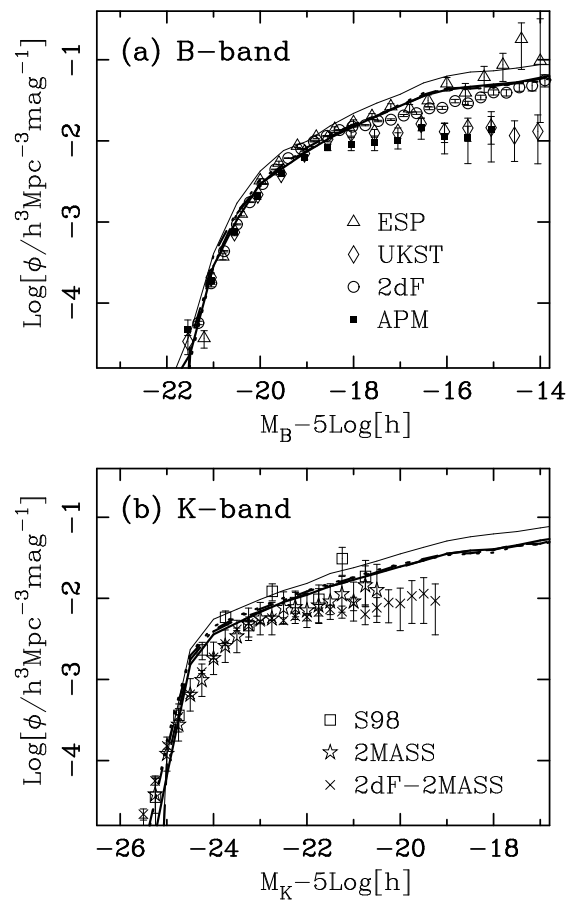}

\caption{Local luminosity functions in the (a) $B$ band and (b) $K$
band.  The thick solid, dashed, dot-dashed, and dotted lines represent
the models of CSF, DSF2, DSF1, and DSF0, respectively, based on the YNY
mass function.  Only for the purpose of comparison here, shown by the
thin solid line is the CSF model based on the original PS mass function.
Symbols with error bars in (a) indicate the observational data from APM
(Loveday et al. 1992, filled squares), ESP (Zucca et al.  1997, open
triangles), Durham/UKST (Ratcliffe et al. 1998, open diamonds), and 2dF
(Folkes et al. 1999, open circles).  Symbols in (b) indicate the data
from Szokoly et al.  (1998, open squares), 2MASS (Kochanek et al. 2000,
open stars), and 2dF combined with 2MASS (Cole et al. 2000, crosses).  }
\label{fig:lf}
\end{figure}

First, the SN feedback-related parameters ($V_{\rm hot}, \alpha_{\rm
hot}$) and merger-related parameter ($f_{\rm mrg}$) are almost uniquely
determined if their values are so chosen as to reproduce the local
luminosity function.  Figure \ref{fig:lf} shows theoretical results for
CSF (solid line), DSF2 (dashed line), DSF1 (dot-dashed line) and DSF0
(dotted line).  As in our previous papers, the SF timescale affects the
local luminosity function only slightly.  Symbols with errorbars
represent observational results from the $B$-band redshift surveys, such
as Automatic Plate Machine \citep[APM;][]{l92}, ESO Slice Project
\citep[ESP;][]{z97}, Durham/United Kingdom Schmidt Telescope
\citep[UKST;][]{r98} and Two-Degree Field \citep[2dF;][]{f99}, and from
the $K$-band redshift surveys given by \citet{s98}, Two Micron All Sky
Survey \citep[2MASS;][]{k01} and 2dF combined with 2MASS \citep{c01}.

In Figure \ref{fig:lf} we see how the mass function of dark halos
affects the luminosity function.  In the same figure we show the same
model as CSF but for the PS mass function represented by the thin solid
lines.  As the PS mass function predicts more dark halos and hence more
galaxies than the YNY mass function except for the largest mass scale.
Thus, we need $V_{\rm hot}=150$ km~s$^{-1}$, less than 280km~s$^{-1}$ in
the previous model, to weaken SN feedback which mainly determines the
scale of the exponential cut-off of the luminosity function
\citep{ntgy01,nytg02}.  We also investigated the effects of power
spectrum of density fluctuations and confirmed that neither a
$\sigma_{8}=1$ model nor a model with BBKS power spectrum without the
baryonic effects is significantly different from our reference model.
In \S\ref{sec:consistency}, we discuss the slight effect that
$\sigma_{8}$ has on high-redshift galaxies.

The effects of photoionization on the luminosity function have been
discussed \citep{cn94, ngs99, bkw00, ng01, s02, tstv02, blbcf02a,
bflbc02b, bfbcl03a}.  There are two ways to suppress the gas cooling by
photoionization.  One effect is that the ultraviolet (UV) background
from quasars and/or young stars prevents hot gas at the outer envelope
from cooling.  Another is that the Jeans mass becomes larger after
cosmic reionization, which means larger $V_{\rm low}$.  It has been
found that both processes lower the faint-end of the luminosity
function.  For example, if we use a large value of $V_{\rm low}\simeq
70$ km~s$^{-1}$ for our model, which corresponds to increasing the Jeans
mass, the number of galaxies at the faint-end of the resultant
luminosity function decreases by about a factor of two, less than that
given by the APM survey at $-20\la M_{B}-5\log(h)\la-17$ (see Figure
\ref{fig:lf}).  In that case we need weaker SN feedback, that is,
smaller $V_{\rm hot}$ and/or smaller $\alpha_{\rm hot}$.  We found,
however, that such a large value of $V_{\rm low}$ does not make dwarf
spheroidals at $M_{B}-5\log(h)\ga -12$ for $V_{\rm low}=70$ km~s$^{-1}$
and at $M_{B}-5\log(h)\ga -10$ for $V_{\rm low}=50$ km~s$^{-1}$.  In
addition, giant galaxies obtain more gas that has not cooled in smaller
halos $V_{\rm circ}\leq V_{\rm low}$, causing the bright-end of the
luminosity function to shift brighter.  Note that this is not effective
for $V_{\rm low}\la 40$ km~s$^{-1}$.  Keeping in mind that such effects
might affect our analysis through the determination of SN
feedback-related parameters, we use $V_{\rm low}=30$ km~s$^{-1}$ in this
paper.

Next, the SFR-related parameter ($\tau_*^0$) is determined by using the
mass fraction of cold gas in spiral galaxies.  The gas fraction 
depends on both the SN feedback-related and SFR-related parameters.
The former parameters determine the gas fraction expelled from galaxies
and the latter the gas fraction that is converted into stars.
Therefore, in advance of determining the SFR-related parameters, the SN
feedback-related parameters must be determined by matching the local
luminosity function.

\begin{figure}
\plotone{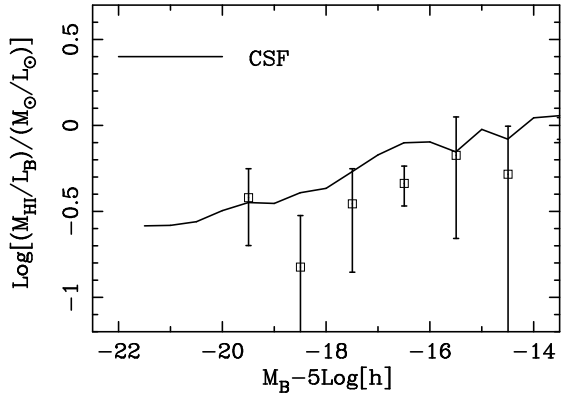}

\caption{Cold gas mass relative to $B$-band luminosity of spiral
galaxies.  The solid line represents the CSF model.  Other SF models,
not shown, give almost the same results with only 0.1 dex difference.  
The open squares indicate the observational data for atomic neutral 
hydrogen taken from Huchtmeier \& Richter (1988).  Since the cold gas 
in the model consists of all species of elements, its mass is multiplied 
by 0.75, i.e., $M_{\rm HI}=0.75M_{\rm cold}$, which corresponds to a total 
fraction of atomic and molecular hydrogen.  Therefore, in comparison with
the theoretical result, the observational data should be regarded as lower 
limits. 
} 
\label{fig:gas}
\end{figure}

Figure \ref{fig:gas} shows the ratio of cold gas mass relative to
$B$-band luminosity of spiral galaxies as a function of their
luminosity.  Theoretical result is shown only for the CSF model by the
solid line.  Other SF models of DSF2, DSF1 and DSF0 provide almost the
same results, and their differences from the CSF are only about 0.1 dex.
We here assume that 75\% of the cold gas in the models is comprised of
hydrogen, i.e., $M_{\rm HI}=0.75M_{\rm cold}$.  \ion{H}{1} data, taken
from \citet{hr88}, are shown by open squares with errorbars.  Since
their data do not include the fraction of H$_{2}$ molecules, they should
be regarded as providing a lower limit to the mass fraction of cold gas.
Adopted values of the parameters in the SF models are tabulated in Table
\ref{tab:astro}.  Slight difference in the values of $\tau_{*}^{0}$
stems from the different redshift-dependence of SF timescale.

\begin{figure}
\plotone{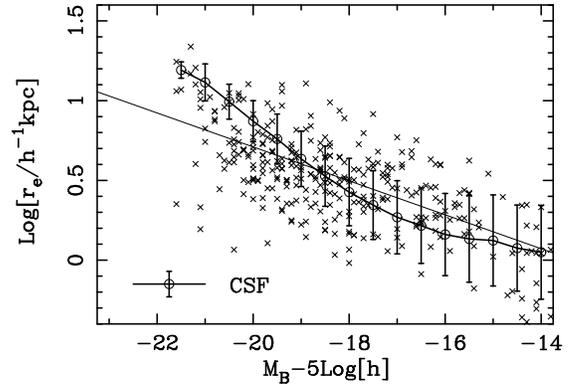}

\caption{Disk size of spiral galaxies.  The thick solid line connecting
open circles shows the theoretical result for effective radii of 
spiral galaxies in the CSF model.  Errorbars on this line is the 
1$\sigma$ scatter in predicted sizes.  Other SF models, not shown, 
give almost the same results with less than 0.1 dex difference.  
The thin line indicates the mean relation given by \citet{ty00} 
for spiral galaxies, based on the observational data (crosses) 
taken from Impey et al. (1996).  
}
\label{fig:r}
\end{figure}

Figure \ref{fig:r} shows the effective disk radii of local spiral
galaxies as a function of their luminosity only for the CSF model.
Other SF models also provide almost the same results as CSF and are not
shown.  Thus, with the use of ($\bar{\lambda}, \sigma_{\lambda}$)=(0.03,
0.5), all the SF models well reproduce the observed disk size-magnitude
relation (thin solid line) compiled by \citet{ty00} based on the data
taken from \citet{isib96}, while showing a slightly steeper slope than
the observed one.

\section{Results}
\subsection{Relationships between Size and Magnitude}
\begin{figure}
\plotone{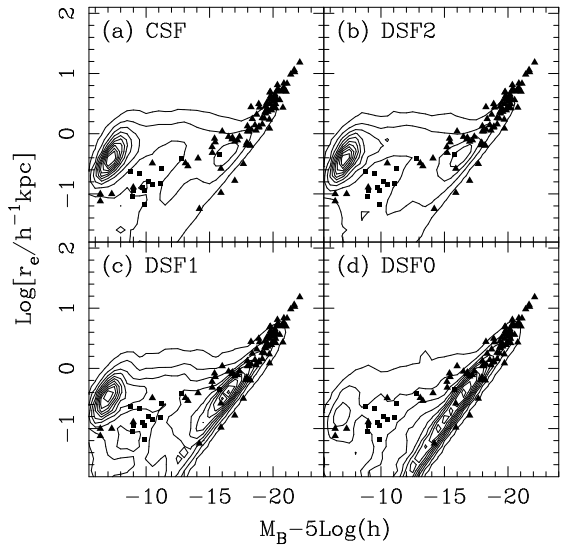}

\caption{Distribution of effective radius for elliptical galaxies in the
 models of (a) CSF, (b) DSF2, (c) DSF1, and (d) DSF0, where the effects
 of dynamical response to starburst-induced gas removal are taken into
 account.  Based on the calculated number of such galaxies per magnitude
 per logarithm of kpc, the levels of contours shown in order from outer
 to inner are 0.02, 0.1, 0.2, 0.3, 0.4, 0.5, 0.6, 0.7, 0.8, and 0.9
 times the largest number of galaxies in grids, respectively.  Symbols
 indicate the observational data given by \citet{bbf92} (triangles) and
 \citet{m98} (squares).  }

\label{fig:rad1}
\end{figure}

First, we examine the size distribution of elliptical galaxies, because
it is directly affected by the dynamical response to gas removal.
Figure \ref{fig:rad1} shows the contour distribution in the effective
radius versus absolute $B$-magnitude diagram.  The levels of contours in
order from outside to inside are 0.02, 0.1, 0.2, 0.3, 0.4, 0.5, 0.6,
0.7, 0.8 and 0.9 times the largest number of galaxies in grids. This way
of description applies to all the contour distributions shown below.
The four panels show the results for the models of CSF, DSF2, DSF1 and
DSF0, respectively, as indicated.  Effective radius $r_{e}$ of model
galaxies are defined by 0.744$r_{b}$, where $r_{b}$ is the
three-dimensional half-mass radius, assuming the de Vaucouleurs profile
\citep{ny03}.  The filled triangles and squares represent the data taken
from \citet{bbf92, bbf93} and \citet{m98}, respectively.  When some of
dwarf spheroidals are listed in both their tables, we use the data in
\citet{bbf92, bbf93}.  Observed elliptical galaxies have two sequences
of dispersed dwarf galaxies with low surface brightness and compact
dwarf galaxies with high surface brightness.

As shown in Figure \ref{fig:sfr}, the CSF model has the longest SF
timescale at high redshift among the four SF models.  This means that
when a major merger occurs in the CSF model, a significant fraction of
gas is removed, so that the dynamical response of dwarf galaxies is
expected to be the largest.  Figure \ref{fig:rad1} clearly shows that
the CSF model ({\it panel a}) predict many dwarf galaxies with very
large size of $r_{e}\sim 10-10^{2}h^{-1}$kpc for $M_{B}-5\log h\ga -10$.
In the range of $-10\ga M_{B}-5\log h\ga-15$, the peak of distribution
is on the sequence of dispersed dwarf ellipticals.  On the other hand,
the DSF0 model ({\it panel d}) has the shortest SF timescale at high
redshift, so that the effects of gas removal are not significant.  Most
of galaxies are distributed along a single power-law sequence,
corresponding to the compact dwarf galaxies.  The slope is determined by
$\alpha_{\rm hot}$ as shown in \citet{ntgy01}, in which they simply
determined the size of elliptical galaxies by $GM_{b}/2V_{b}^{2}$.  The
behavior of SF timescale indicates that the models of DSF1 and DSF2 are
intermediate between CSF and DSF0.  This suggests that the SF timescale
should be constant independent of redshift or very weakly dependent on
redshift, but not proportional to the dynamical timescale of galaxies.

\begin{figure}
\plotone{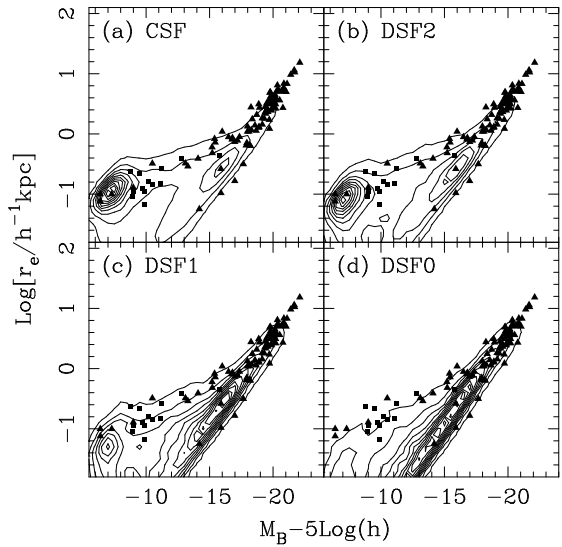}

\caption{Distribution of effective radius for elliptical galaxies. Same
as Figure \ref{fig:rad1}, but for the SF models without the effects of
dynamical response to starburst-induced gas removal.  }

\label{fig:rad2}
\end{figure}

It should be noted that the gas removal is not a unique mechanism to make 
large ellipticals.  If a galaxy just after major
merger keeps a large fraction of gas, its size becomes larger than that 
of gas-poor system with the same stellar mass because the size is proportional
to the total mass of stars and cold gas.
This can be seen directly if we do not take into account the effects of
dynamical response to gas removal.  Figure \ref{fig:rad2} shows
the same result as Figure \ref{fig:rad1} except without the effects of
dynamical response.  The overall shapes are similar, independent of SF 
timescale, because the effects of dynamical response are not taken into 
account.  However the peak locations in Figure \ref{fig:rad2} are 
different from those in Figure \ref{fig:rad1}.  This is because the mass
of cold gas just after major merger determines its size as mentioned above 
and therefore the size depends on the SF timescale.  In other words, the 
CSF model has a larger fraction of cold gas and hence a larger galactic 
mass at high redshift, compared with the DSF models.
This situation without the effects of dynamical response 
corresponds to the case of {\it dominant dark halo} in \citet{ds86},
in which the size and velocity dispersion do not change during the gas 
removal.  We therefore conclude that in realistic situations the
dark matter is not always dominant in the gravitational potential.  Note
that while the $r_e - M_B$ relation is reproduced even without the 
effects of the dynamical response, as will shown in the next subsection, 
the Faber-Jackson relation is not reproduced unless the effects are
properly taken into account (see Figures \ref{fig:fj1} and \ref{fig:fj2}).

\begin{figure}
\plotone{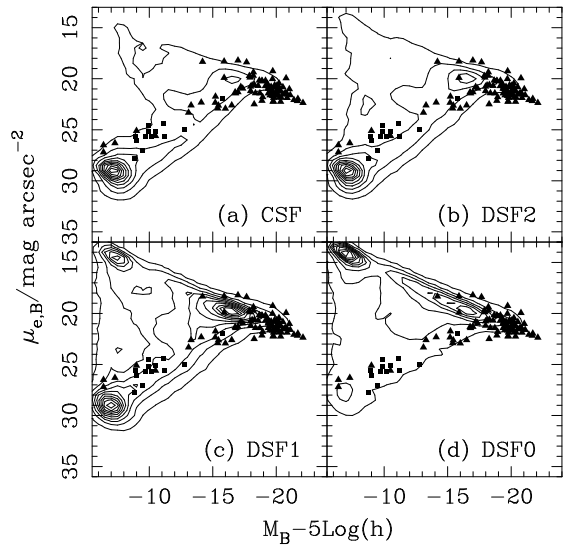}

\caption{Distribution of $B$-band surface brightness within the
effective radius for elliptical galaxies in the models of (a) CSF, (b)
DSF2, (c) DSF1, and (d) DSF0, where the effects of dynamical response to
starburst-induced gas removal are taken into account. Contours and
symbols have the same meanings as in Figure \ref{fig:rad1}.  }

\label{fig:sb1}
\end{figure}

In addition to size and magnitude, surface brightness is also an
important observable quantity.  Two sequences of dispersed and compact
dwarf galaxies are prominent in the surface brightness versus absolute
$B$-magnitude diagram, as shown in Figure \ref{fig:sb1}.  Surface
brightness of model galaxies is defined as the average brightness in the
area encircled by effective radius.  For giant ellipticals, surface
brightness is predicted to become brighter towards brighter magnitude
with shallow slope, converging to $\mu_{e,B}\simeq 22$, as observed.
For $M_{B}-5\log h\ga -18$, surface brightness widely spreads.  This
magnitude corresponds to $V_{\rm hot}$, that is, the magnitude at which
$\beta$ becomes larger than unity.  The CSF and DSF2 models make the
widely spread distribution which reproduces two sequences simultaneously
and thus are likely, while the DSF0 makes one sequence of compact
ellipticals only.

\begin{figure*}
\plotone{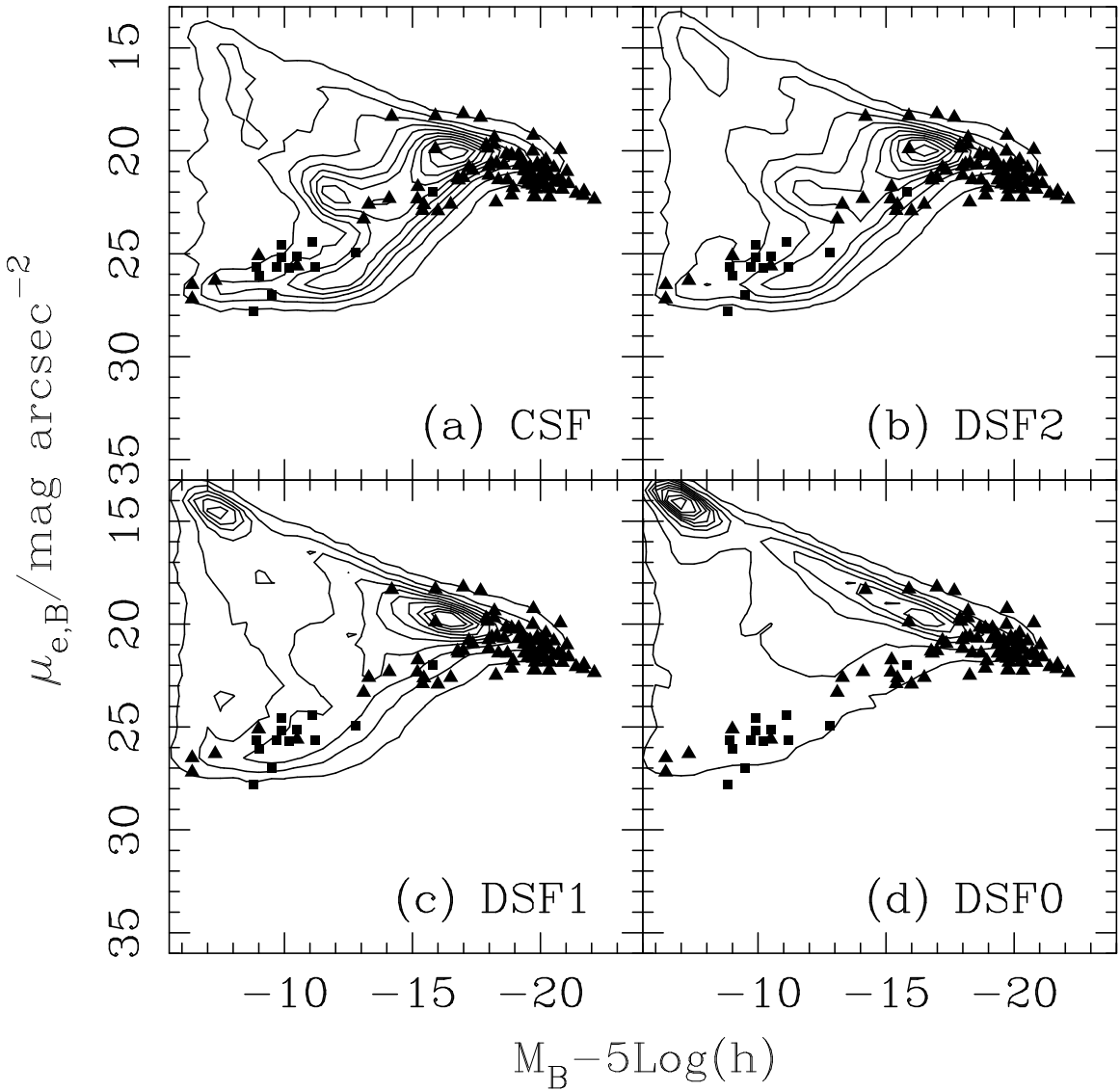}

\caption{Distribution of $B$-band surface brightness within the
effective radius for elliptical galaxies. Same as Figure \ref{fig:sb1},
but for only galaxies of high surface brightness with $\mu_{e,B}\leq
26.5$ in the models as well as the data.  }

\label{fig:sb3}
\end{figure*}

\begin{figure*}
\plotone{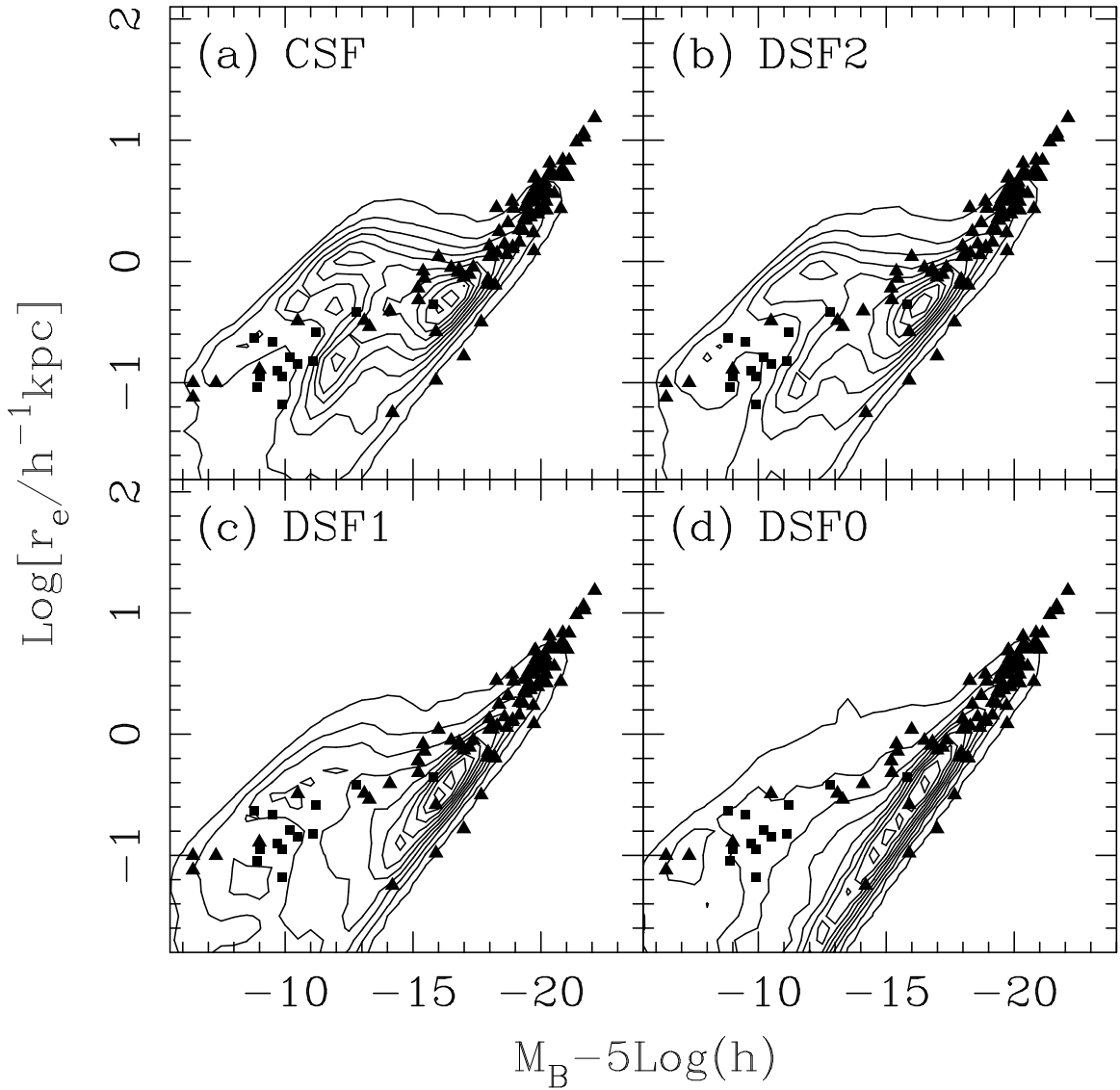}

\caption{Distribution of effective radius for elliptical galaxies.  Same
as Figure \ref{fig:rad1}, but for only galaxies of high surface
brightness with $\mu_{e,B}\leq 26.5$ in the models as well as the data.  
}

\label{fig:rad3}
\end{figure*}

In all the SF models there are many galaxies with very low surface
brightness for $M_{B}-5\log h\ga-15$.  Actually such galaxies cannot be
detected unless the detection threshold of surface brightness is faint
enough in galaxy survey observations.  With a cutoff of $\mu_{e,B}=26.5$
taken into account in both the model and observed data, we show the
surface brightness versus absolute $B$-magnitude diagram in Figure
\ref{fig:sb3} and the effective radius versus absolute $B$-magnitude
diagram in Figure \ref{fig:rad3}.  For the models of CSF and DSF2 the
resultant distributions in both of these diagrams are consistent with
the data.

\begin{figure*}
\plotone{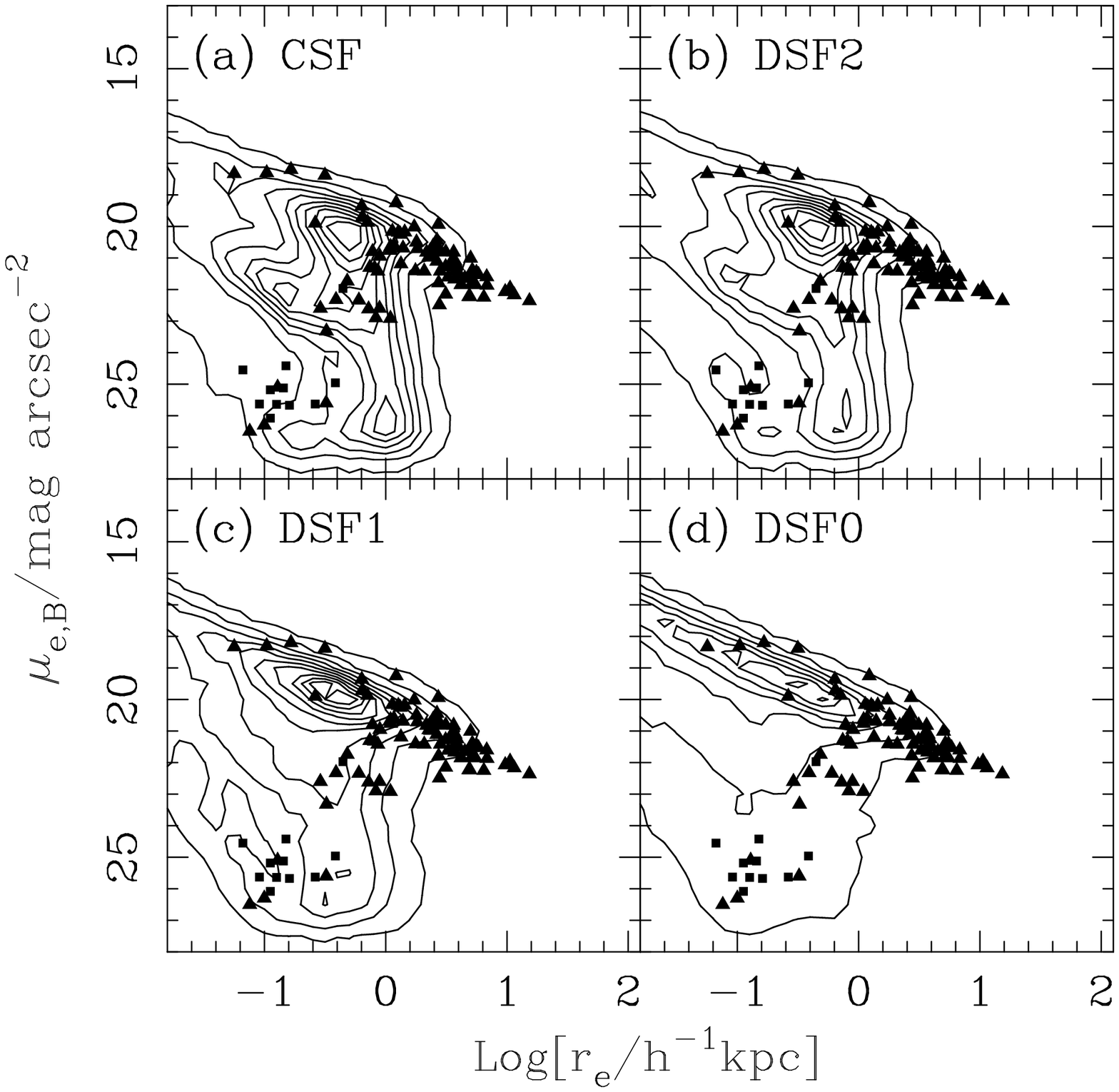}

\caption{Kormendy relation for elliptical galaxies with $\mu_{e,B}\leq
26.5$ in the models of (a) CSF, (b) DSF2, (c) DSF1, and (d) DSF0, where
the effects of dynamical response to starburst-induced gas removal are
taken into account.  Contours and symbols have the same meanings as in
Figure \ref{fig:sb3}.  }

\label{fig:kor}
\end{figure*}

As another combination of observable quantities, the surface brightness
versus size relation is often referred to as the Kormendy relation
\citep{k77}.  Figure \ref{fig:kor} shows the theoretical distribution of
elliptical galaxies with $\mu_{e,B}\leq 26.5$ in the surface brightness
versus effective radius diagram.  Extended distribution toward low
surface brightness is clearly seen for the models of CSF and DSF2.

\subsection{Faber-Jackson Relation}
Velocity dispersion of elliptical galaxies is also an independent
dynamical observable.  If the dark matter dominates the baryonic 
matter, the velocity dispersion of galaxies reflects that of 
their host dark halos.  On the other hand, in the case
of negligible dark matter, the velocity dispersion is substantially
affected by dynamical response to gas removal in proportion to the
mass fraction of removed gas.  Thus the velocity dispersion versus
absolute magnitude relation for elliptical galaxies, 
often called Faber-Jackson (FJ) relation, provides another strong 
constraint on galaxy formation \citep{fj76}.

\begin{figure}
\plotone{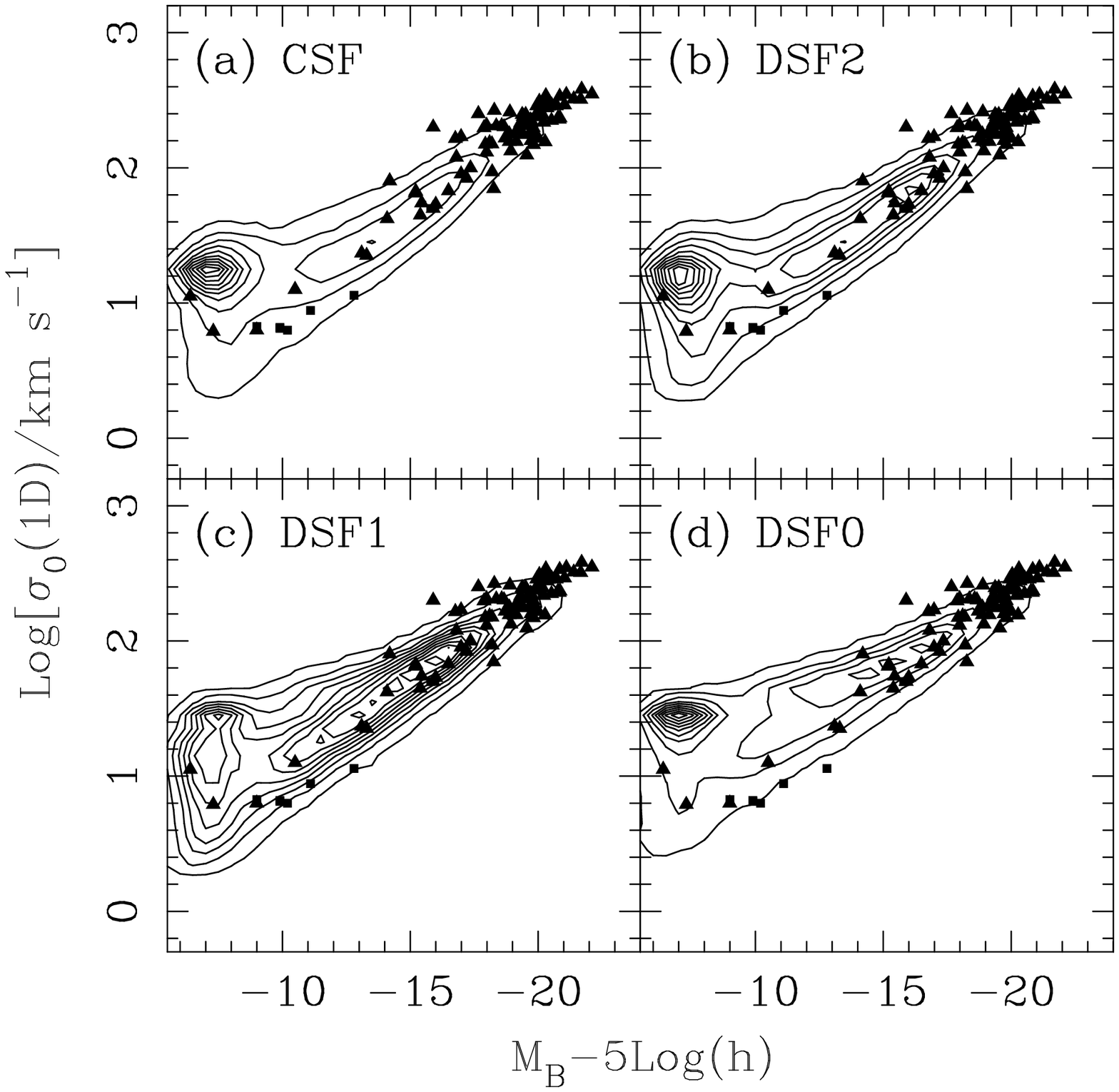}

\caption{Faber-Jackson relation for elliptical galaxies in the models of
(a) CSF, (b) DSF2, (c) DSF1, and (d) DSF0, where the effects of
dynamical response to starburst-induced gas removal are taken into
account.  One-dimensional central velocity dispersion $\sigma_{0}(\rm
1D)$ for model galaxies is estimated from $V_{b}/\sqrt{3}$ after
increased to the central value by a factor of $\sqrt{2}$ according to
the de Vaucouleurs-like profile. Contours and symbols have the same
meanings as in Figure \ref{fig:rad1}.}

\label{fig:fj1}
\end{figure}

Figure \ref{fig:fj1} shows the velocity dispersion versus absolute
$B$-magnitude diagram for the models of CSF, DSF2, DSF1 and DSF0.
Velocity dispersion of model galaxies is assumed to be isotropic and is
converted to one-dimensional central dispersion by $\sigma_{0}(\rm
1D)=V_{\rm b}/\sqrt{3}$ after increased to the central value by a factor
of $\sqrt{2}$ according to the de Vaucouleurs-like profile.  For the
DSF0 model, galaxies with velocity dispersion less than 10km s$^{-1}$ is
scarcely distributed.  Note that the cutoff circular velocity $V_{\rm
low}$, above which dark halos are identified as isolated objects, is
30km s$^{-1}$ nearly corresponding to the Jeans scale for collapse after
cosmic reionization \citep{og96}.  This makes the sequence that
converges to $\sigma_{0}(\rm 1D)=30/\sqrt{3}\simeq 17$ km s$^{-1}$.  For
the DSF0 model, we see only such sequence because the smallest fraction
of gas at major merger in this model gives very weak effects of
dynamical response to gas removal.  It should be noted that this
tendency hardly depends on $V_{\rm low}$ because the SN feedback is very
efficient for dwarf galaxies.  In contrast, other SF models can
reproduce the low velocity dispersion observed for local dwarf
spheroidals for $M_{B}-5\log h\ga-15$.

\begin{figure*}
\plotone{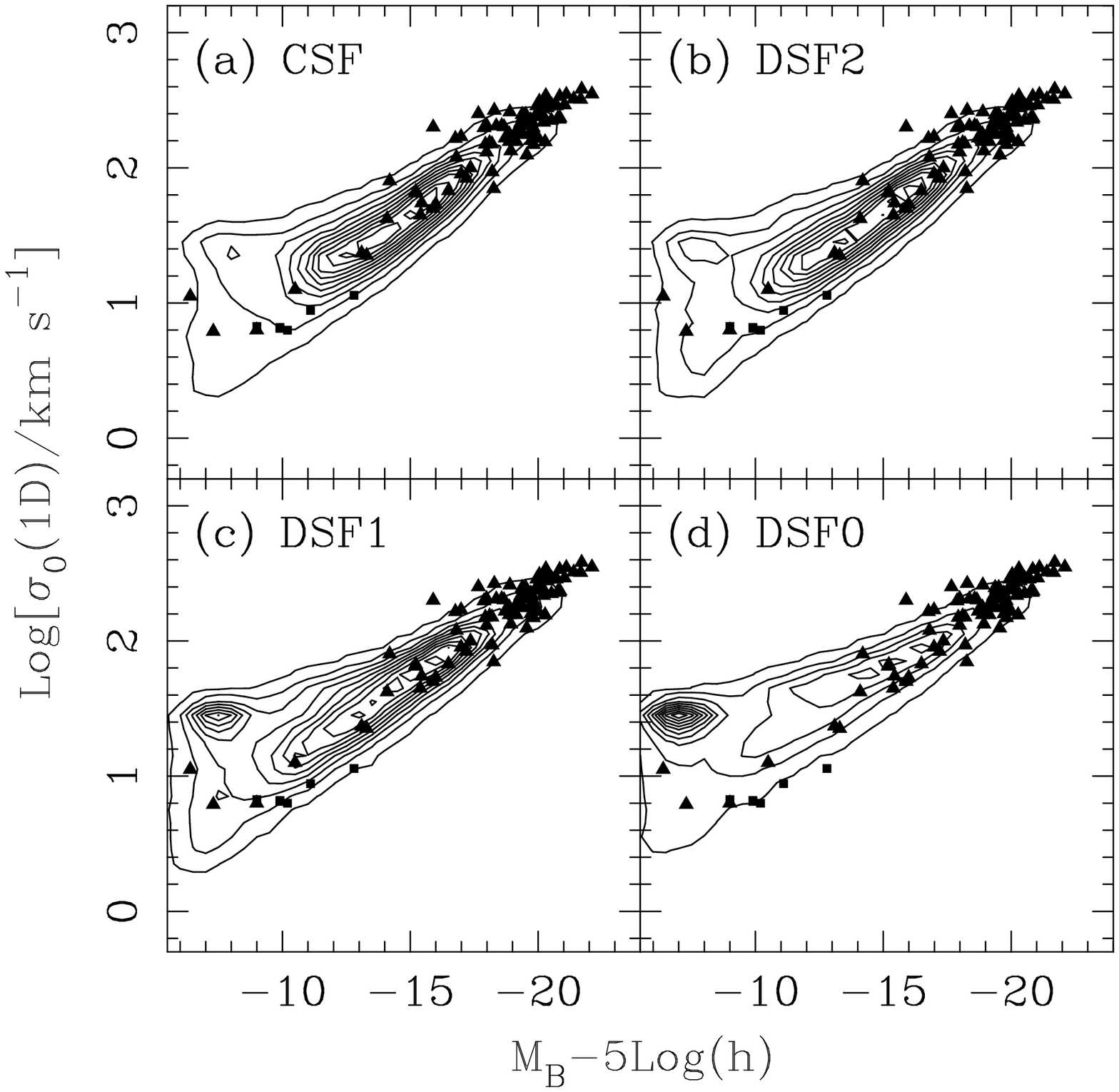}

\caption{Faber-Jackson relation for elliptical galaxies.  Same as Figure
\ref{fig:fj1}, but for only galaxies of high surface brightness with
$\mu_{e,B}\leq 26.5$ in the models as well as the data.  
}

\label{fig:fj3}
\end{figure*}

It is possible that dwarf galaxies with very low velocity dispersion
would remain undetected from actual observations, because such galaxies
are expected to have very low surface brightness as a result of
dynamical expansion by gas removal.  Figure \ref{fig:fj3} therefore
shows the FJ relation after excluding galaxies of lower surface
brightness with $\mu_{e,B}\geq 26.5$.  Most of dwarf galaxies of low
circular velocity are excluded from the CSF model and their distribution
is in apparent disagreement with the data.  This situation is a little
improved for the models of DSF1 and DSF2.  Figure \ref{fig:fj3} clearly
indicates the difficulty of making dwarf galaxies of low circular
velocity and high surface brightness.  In other words, the FJ relation
strongly constrains the SF timescale requiring the mild or negligible
redshift dependence, when compared with the dynamical timescale.

\begin{figure}
\plotone{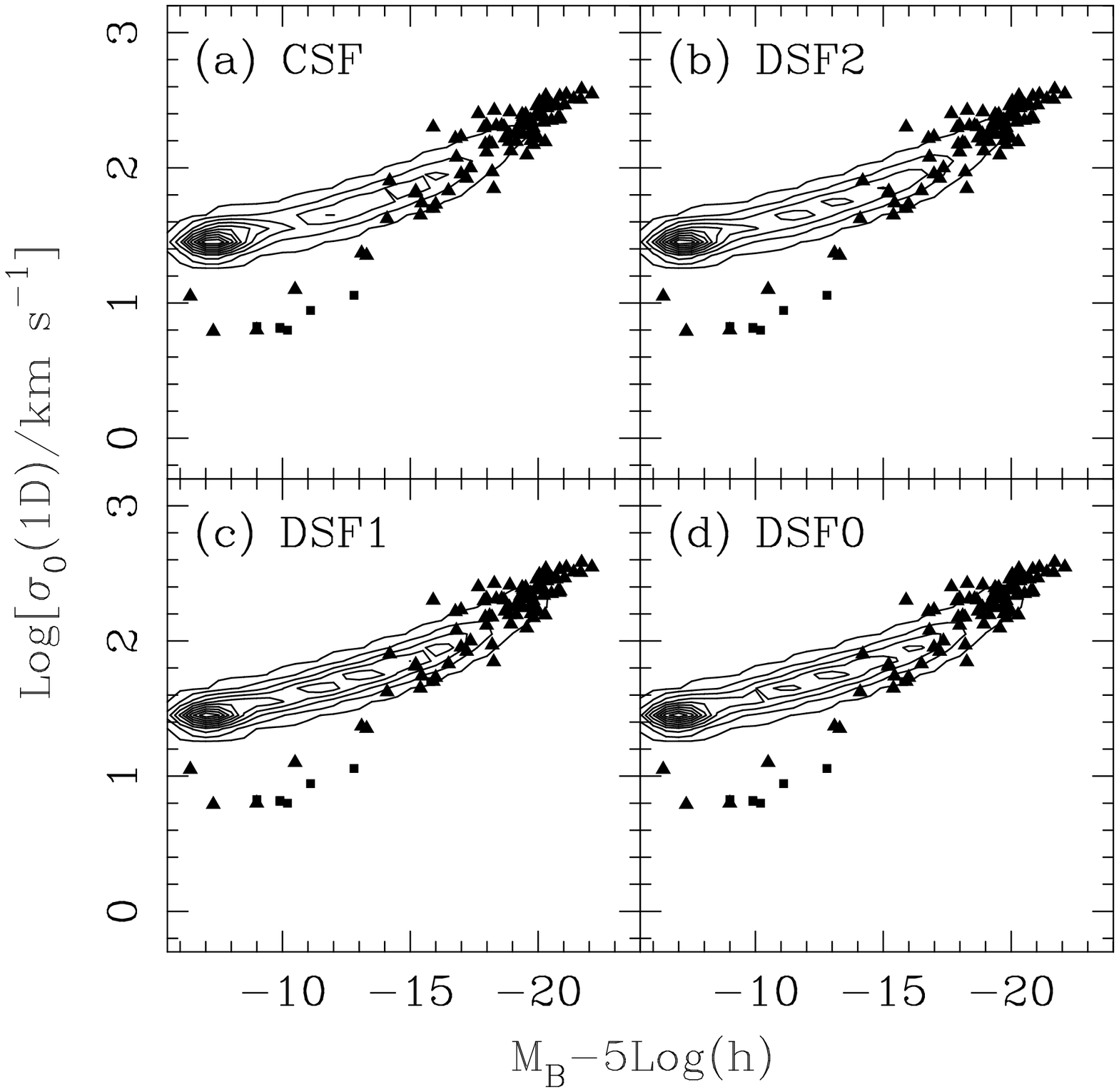}

\caption{Faber-Jackson relation for elliptical galaxies.  Same as Figure
\ref{fig:fj1}, but for the SF models without the effects of dynamical
response to starburst-induced gas removal.  }

\label{fig:fj2}
\end{figure}

Here we examine the effects of the dynamical response on velocity
dispersion.  Figure \ref{fig:fj2} shows the velocity dispersion versus
absolute $B$-magnitude diagram without the effects of dynamical 
response as in Figure \ref{fig:rad2}.  Evidently, all four SF models 
cannot reproduce the dwarf galaxies of low circular velocity and give 
almost the same distribution on this diagram independent of SF timescale.
This hilights the importance of dynamical response for the formation of 
dwarf galaxies.  \citet{kc98} predicted the FJ relation 
in their SAM with no dynamical response taken into account and claimed to
find good agreement with observations.  Their result is, however, limited 
only to $M_{B}\la -18$, where such dynamical response has no significant
effect.  Thus our SAM analysis is the first that has reproduced the 
observed velocity dispersion even for dwarf galaxies.

\subsection{Mass-to-Light Ratio}
\begin{figure}
\plotone{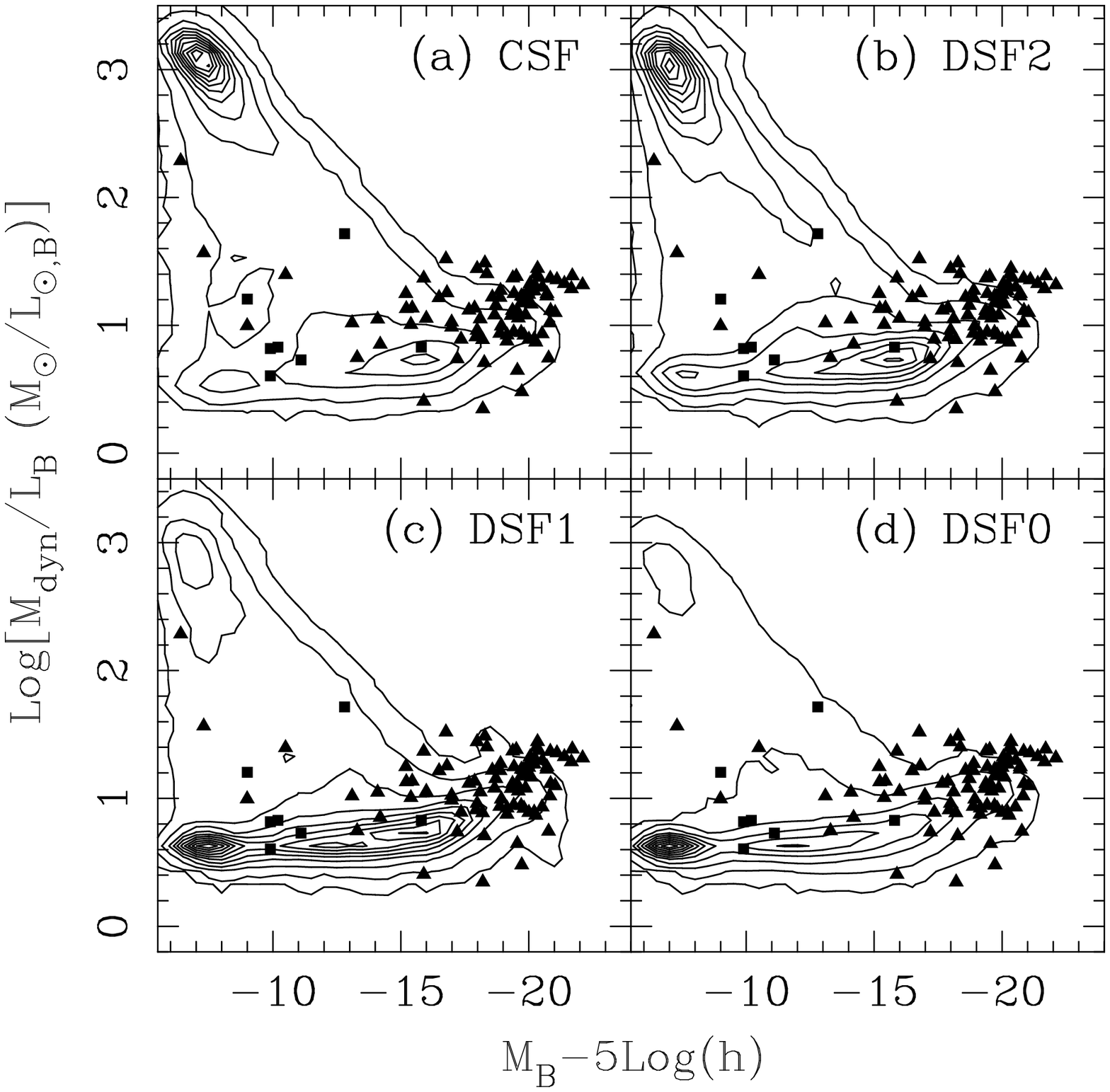}

\caption{Distribution of $B$-band mass-to-light ratio for elliptical
galaxies for the models of (a) CSF, (b) DSF2, (c) DSF1, and (d) DSF0,
where the effects of dynamical response to starburst-induced gas removal
are taken into account.  Contours and symbols have the same meanings as
in Figure \ref{fig:fj1}, but for the outermost contour level added showing
10$^{-3}$ times the largest number of galaxies in grids.  }

\label{fig:ml1}
\end{figure}

The mass, which is used to estimate the mass-to-light ratio, is the {\it
dynamical mass} of $M_{\rm dyn}\propto r_{b}\sigma^{2}/G$.  In our SAM,
the size $r_{b}$ and the velocity dispersion $\sigma$ are determined by
the amount of baryonic matter that escapes out of the system against the
underlying dark matter potential. In this paper we define the dynamical
mass by $M_{\rm dyn}=2r_{b}V_{b}^{2}/G$.  Figure \ref{fig:ml1} shows the
mass-to-light ratio of elliptical galaxies as a function of absolute
$B$-magnitude for the four SF models.  

Our SAM well reproduces the observed mass-to-light ratio for giant
ellipticals, and at least qualitatively the mass-to-light ratio for
dwarf spheroidals that increases towards faint magnitude.  As for giant
ellipticals, observed data still have a large scatter.  This might be
caused by an uncertainty in estimating dynamical mass owing to
anisotropic kinematics and rotation.

\begin{figure}
\plotone{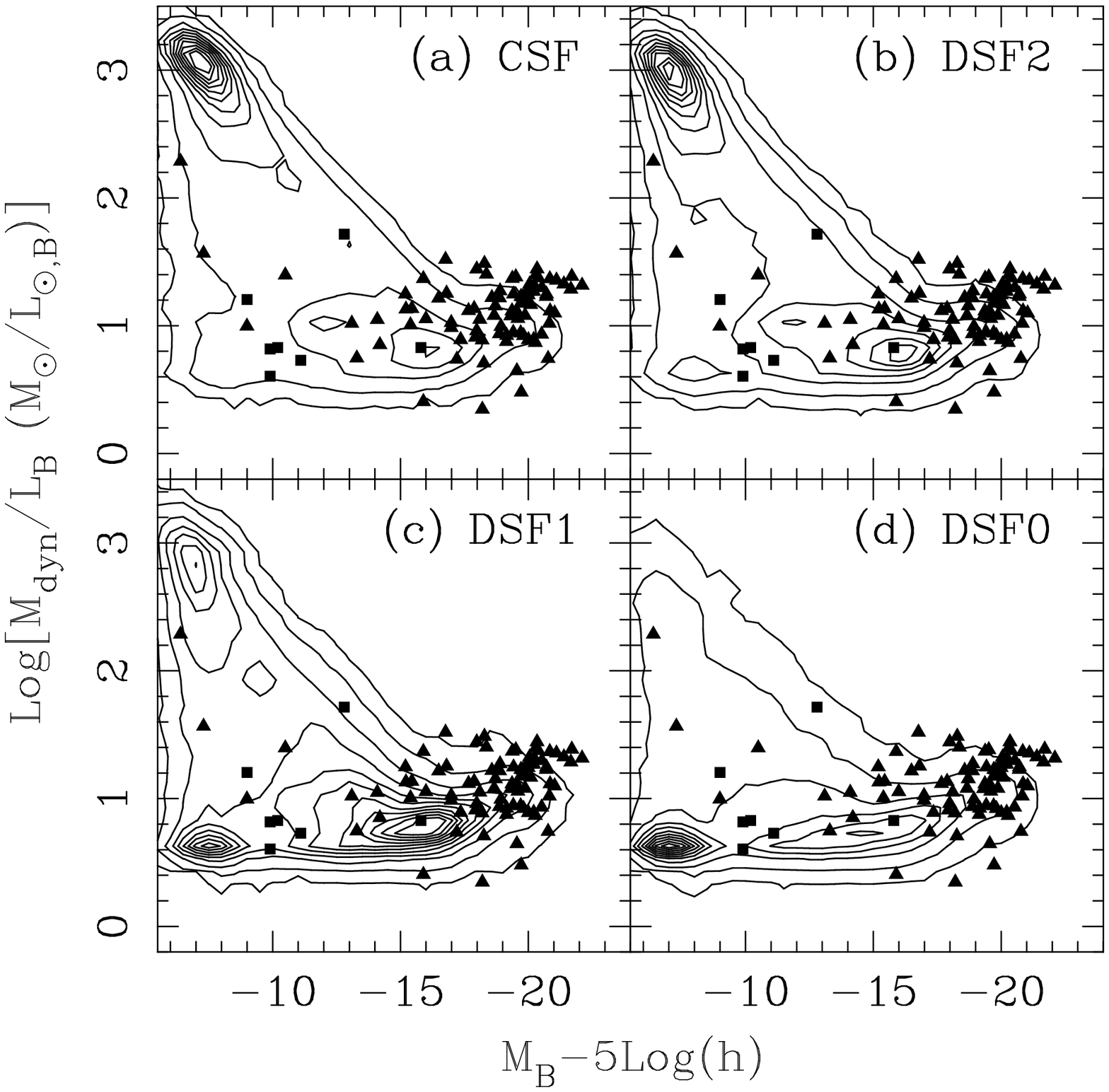}

\caption{Distribution of $B$-band mass-to-light ratio for elliptical
galaxies. Same as Figure \ref{fig:ml1}, but for the SF models without
effects of dynamical response to starburst-induced gas removal.  }

\label{fig:ml2}
\end{figure}

Such trend of the mass-to-light ratio for dwarf spheroidals is not
caused by the dynamical response alone.  Figure \ref{fig:ml2} shows the
mass-to-light ratio with no dynamical response taken into account, which
turns out to be very similar to or slightly higher than that with
dynamical response in Figure \ref{fig:ml1}.  Therefore, considerably
high values up to $M/L_B\sim 10^{3}$ are caused by the dominant dark
halo in which the size and velocity dispersion are kept unchanged during
the starburst.  Since the luminosity $L_{B}$ is proportional to the
final mass $M_{f}$, it follows that $M/L_{B}\propto
M_{i}/M_{f}\propto\beta$ in the limit of $\beta\gg 1$, where $\beta$
measures the strength of the SN feedback in equation (\ref{eqn:beta}).
Using a relation $M\propto V_{\rm circ}^{3}$ from the spherical collapse
model \citep{t69, gg72}, we obtain
\begin{equation}
\frac{M}{L_{B}}\propto\beta\propto \sigma_{i}^{-\alpha_{\rm hot}}\propto
 M_{f}^{-\alpha_{\rm hot}/(3+\alpha_{\rm hot})}.
\end{equation}
If we adopt $\alpha_{\rm hot}=4$, then $M/L_{B}\propto  M_{f}^{-4/7}
\propto 10^{8M_{B}/35}$.  The slope of this relation explains the 
result in Figure \ref{fig:ml2}.  Therefore, the smaller mass fraction 
of dark matter lowers the mass-to-light ratio \citep{ds86, ya87}.  
In apparent contrast to this, the DSF0 model scarcely forms galaxies of
such high mass-to-light ratio.  This is because the size is
determined by the baryonic mass, $r_{e}\propto GM_{b}/V_{b}^{2}$, and
because the removed gas is negligible due to the short SF timescale
at high redshift.

\begin{figure*}
\plotone{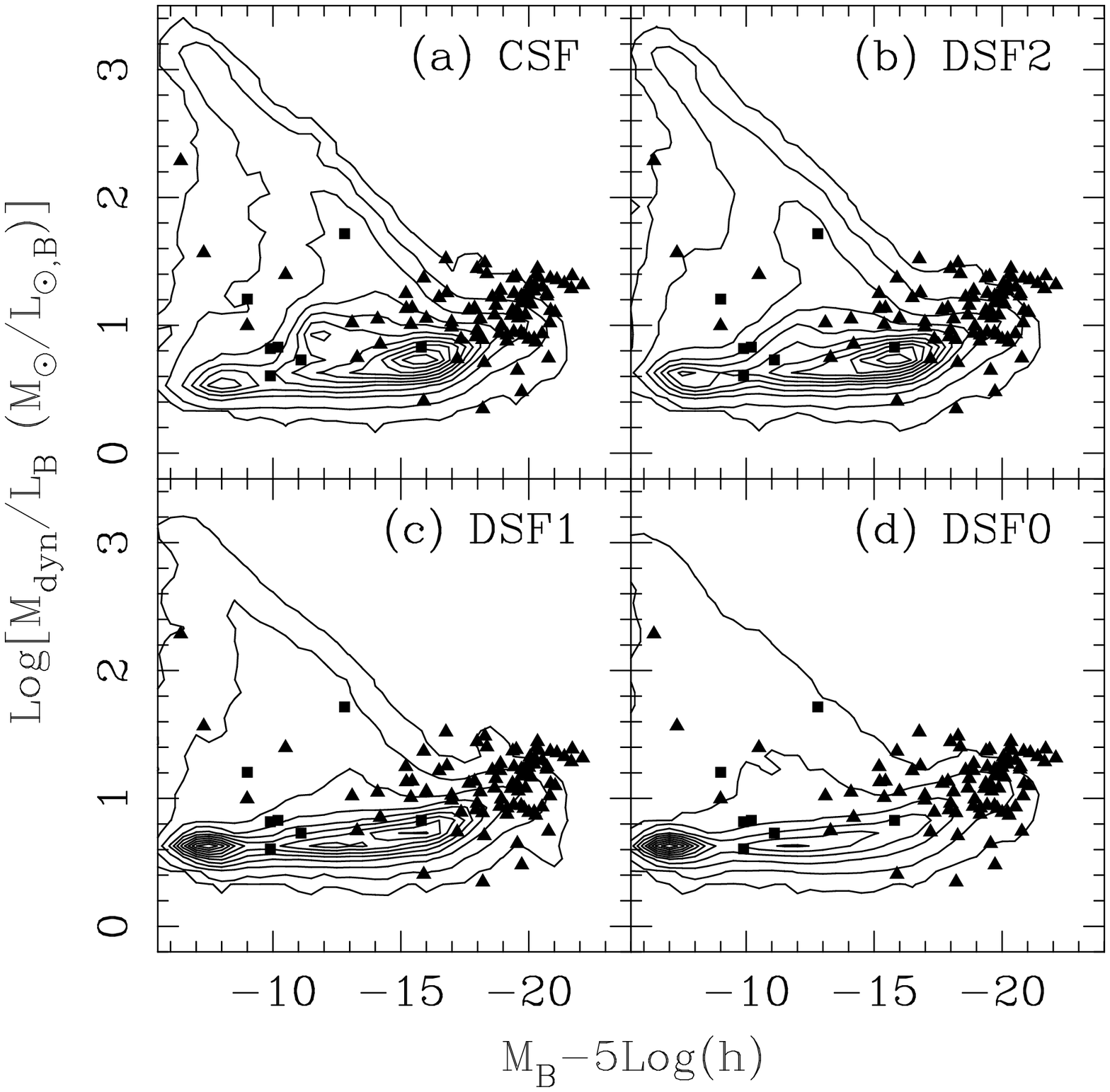}

\caption{Distribution of $B$-band mass-to-light ratio for elliptical
galaxies. Same as Figure \ref{fig:ml1}, but for only galaxies of high
surface brightness with $\mu_{e,B}\leq 26.5$ in the models as well as
the data.  }

\label{fig:ml3}
\end{figure*}

By excluding galaxies of lower surface brightness with 
$\mu_{e,B}\geq 26.5$, we obtain gross agreement with the observed 
mass-to-light ratio in Figure \ref{fig:ml3}.

\subsection{Metallicity-Magnitude Relation}\label{sec:metallicity}
\begin{figure*}
\plotone{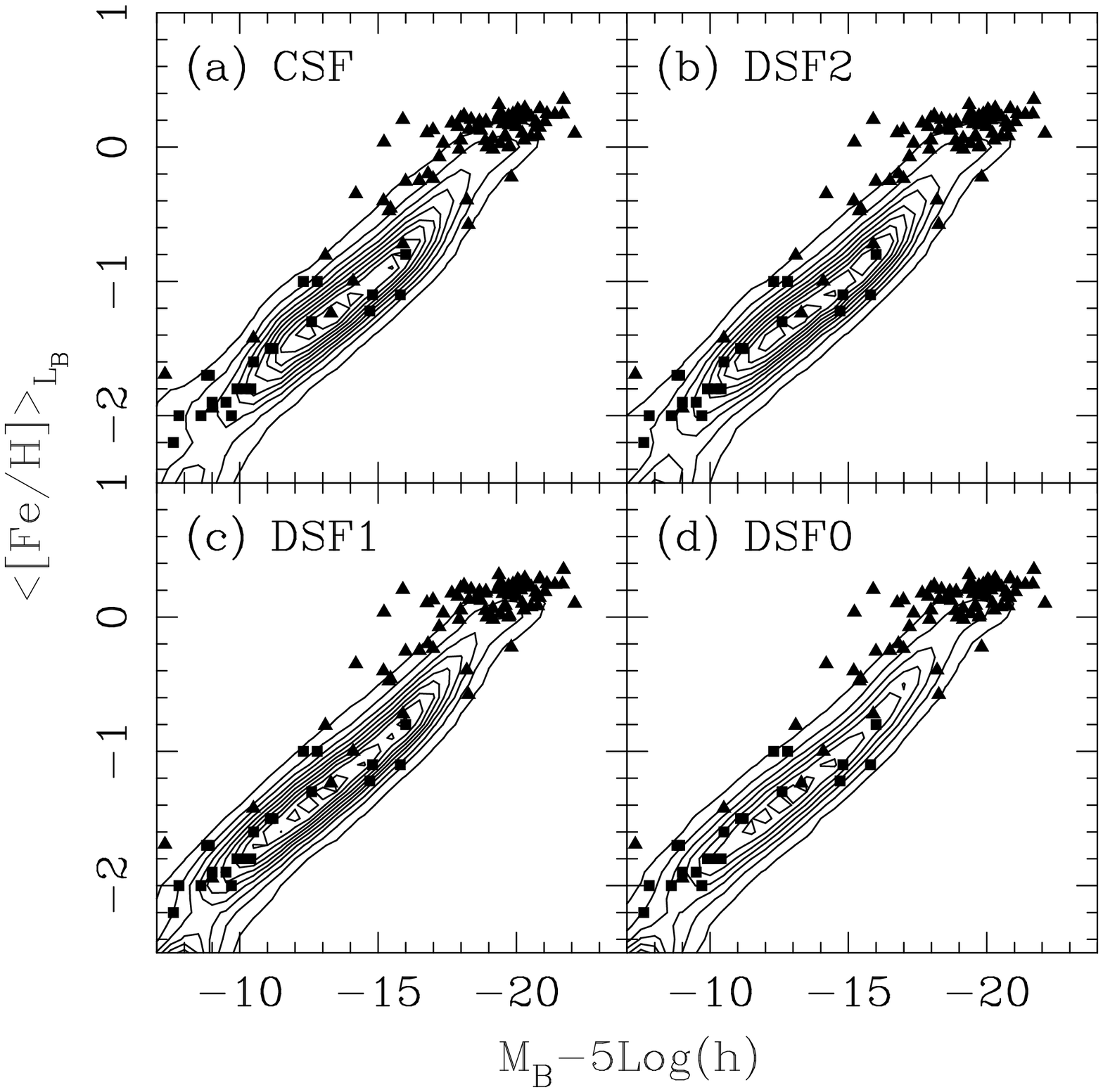}

\caption{Distribution of mean stellar metallicity for elliptical
galaxies in the models of (a) CSF, (b) DSF2, (c) DSF1, and (d) DSF0,
where the effects of dynamical response to starburst-induced gas removal
are taken into account. Contours and symbols have the same meanings as
in Figure \ref{fig:sb3}.  Only galaxies of high surface brightness with
$\mu_{e,B}\leq 26.5$ are shown, because the theoretical results in this
figure are not sensitive to this threshold.  The logarithmic iron
abundance [Fe/H] is used for the data compiled by \citet{m98} and those
translated from Mg$_{2}$ index by \citet{bbf93}, whereas the
$L_{B}$-weighted average of logarithmic metal abundances of stars
$\langle\log(Z_{*}/Z_{\odot})\rangle_{L_{B}}$ for model galaxies.  }

\label{fig:metal3}
\end{figure*}

Figure \ref{fig:metal3} shows the mean stellar metallicity as a function
of absolute $B$-magnitude for elliptical galaxies.  The vertical axis
represents the logarithmic iron abundance [Fe/H].  Shown by symbols are
the data compiled by \citet{m98} and those translated from Mg$_{2}$
index for those taken by \citet{bbf93}, whereas by contours the
$L_B$-weighted average of logarithmic metal abundances of stars
$\langle\log(Z_{*}/Z_{\odot})\rangle_{L_{B}}$ for the theoretical
prediction.  Since the theoretical results are almost independent of
whether galaxies of low surface brightness are excluded, we only show
the metallicity distribution excluding such galaxies with $\mu_{e,B}\geq
26.5$.  We do not show the metallicity distribution for spirals, which
is similar to that for ellipticals.

All the SF models well reproduce the observed tight relation between 
metallicity and absolute magnitude.  This means that such relation 
is not affected by the redshift dependence of SF timescale, because 
we adopt the SN feedback in the same way irrespective of either 
continuous or burst formation of stars in galaxies.  With strong 
SN feedback with $\alpha_{\rm hot}\simeq 4$, our SAM succeeds to 
explain the observed low metallicity [Fe/H]$\sim -2$ at 
$M_{B}-5\log h\simeq -10$, in spite of using a rather high value of
metallicity yield $y=2Z_{\odot}$.  Note that if $\alpha_{\rm hot}=2$, 
we obtain about a factor of three higher metallicity at that 
magnitude.  This is consistent with the conclusion by \citet{c00} 
that with $\alpha_{\rm hot}=2$ the metallicities of their dwarf 
galaxies are systematically higher than the data, while such 
predicted values reside in a range of large observed scatter.

\section{Consistency with Other Observations}\label{sec:consistency}
In this section, we show the whole aspects of our SAM to check the
consistency with other local and high-redshift observations.  This will
clarify the limitations of present SAM analyses and physical processes 
for which further investigation is required.  
In the first three subsections (\S\S5.1-5.3), high-redshift properties 
of model galaxies are examined by comparing the observed galaxies in 
the Hubble Deep Field \citep[HDF,][]{w96} and the Subaru Deep Field 
\citep[SDF,][]{m01}.
Then, the Tully-Fisher relation (TFR) for local spiral galaxies (\S5.4)
and the color-magnitude relation (CMR) for cluster elliptical galaxies 
(\S5.5) are discussed.

\subsection{Faint Galaxy Number Counts}\label{sec:counts}
Counting galaxies as a function of apparent magnitude is one of the
most important observable quantities for constraining the geometry of
the universe and the evolution of galaxies \citep[e.g.,][]{yt88}.  We
already showed that our SAM can simultaneously reproduce the UV/optical
galaxy counts in the HDF \citep{ntgy01} and the near-infrared galaxy
counts in the SDF \citep{nytg02}.  These previous works demonstrated
that inclusion of the selection effects arising from the cosmological
dimming of surface brightness of high-redshift galaxies derived by
\citet{y93} is essential, because the number count of galaxies is
obtained by summing up the product of luminosity function and
cosmological volume element to the accessible maximal redshift above
which galaxies have the surface brightness fainter than the threshold
and thus are not detected. Details of the method to estimate such
selection effects are found in \citet{y93}, \citet{ty00}, \citet{t01}
and \citet{ntgy01, nytg02}.

\begin{figure*}
\plotone{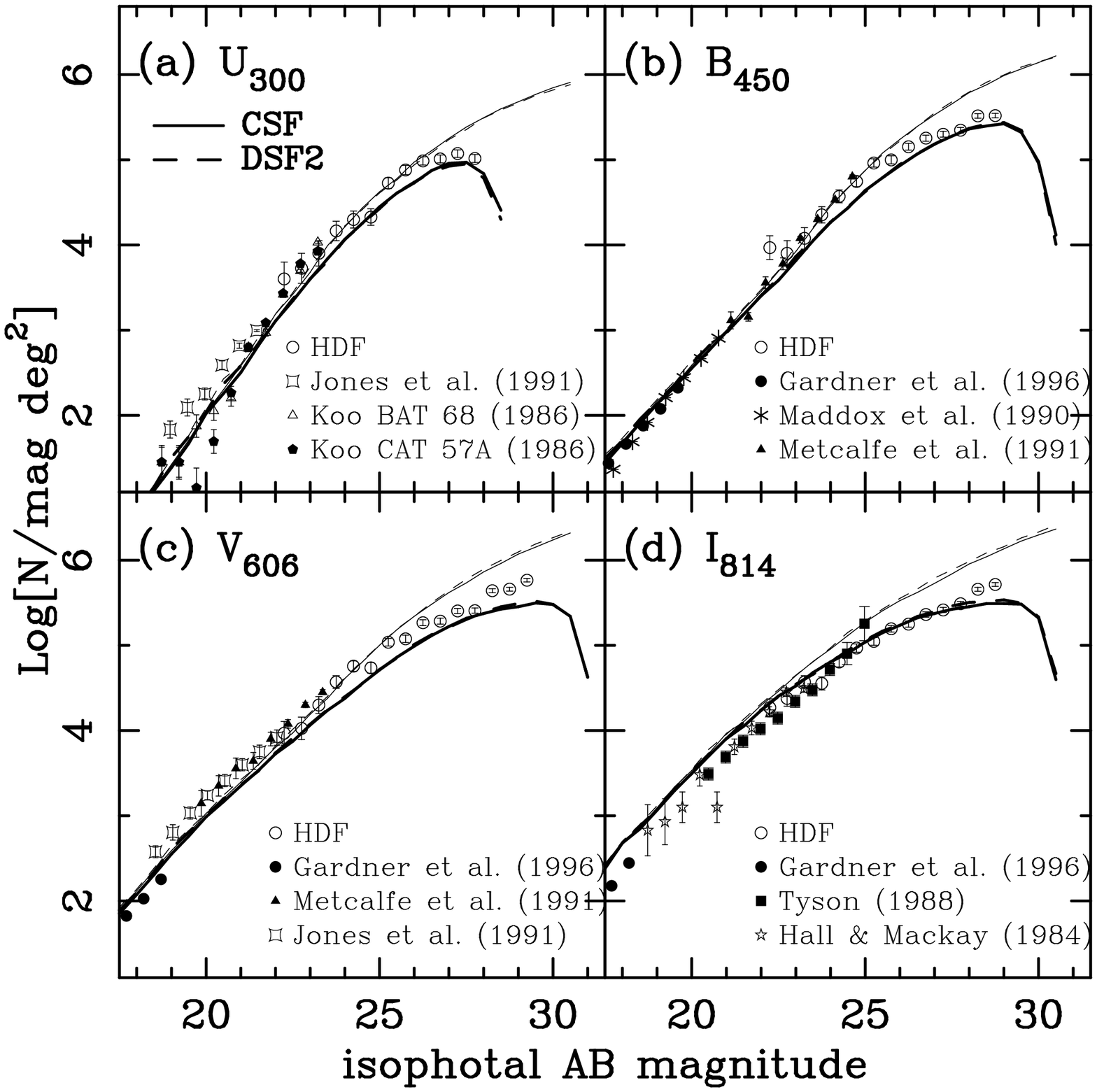}

\caption{Galaxy number counts in the (a) $U_{300}$, (b) $B_{450}$, (c)
$V_{606}$ and (d) $I_{814}$ bands, respectively.  The thick and thin
lines represent the SF models with and without the selection effects,
respectively, arising from the same detection threshold of surface
brightness of galaxies as employed in the HDF survey. For each of the
cases, the solid and dashed lines represent the models of CSF and DSF2,
respectively, where the effects of dynamical response to
starburst-induced gas removal are taken into account. Open circles with
errorbars indicate the observed HDF counts.  Other symbols are
ground-based observed counts.  }

\label{fig:hdf}
\end{figure*}

In this paper the YNY mass function of dark halos (\S\S\ref{sec:mh}) is
used instead of the PS, and the dynamical response to gas removal is a
novel ingredient in the analysis.  With these modifications we reexamine
the galaxy counts again.  Figure \ref{fig:hdf} shows the UV/optical
galaxy counts in the HDF ($U_{300}, B_{450}, V_{606}$ and $I_{814}$ in
the AB system).  For reference, other ground-based observed counts are
also plotted.  The thick solid and dashed lines show the predictions of
CSF and DSF2, respectively, taking into account the absorption by
internal dust \citep{ty00} and by intervening \ion{H}{1} clouds
\citep{yp94}, and the selection effects due to cosmological dimming of
surface brightness \citep{y93}.  We adopt $\rho=1$ in equation
(\ref{eqn:sizerho}), which determines the disk size at high redshift.
The thin lines represent the same models, except without the selection
effects.  We do not show the models of DSF1 and DSF0, because they are
much the same as CSF and DSF2 with only small difference comparable to
observational errors.  These models only slightly underpredict the
$U_{300}$ count and overpredict the $I_{814}$ count, as pointed out by
\citet{ntgy01}.

\begin{figure}
\plotone{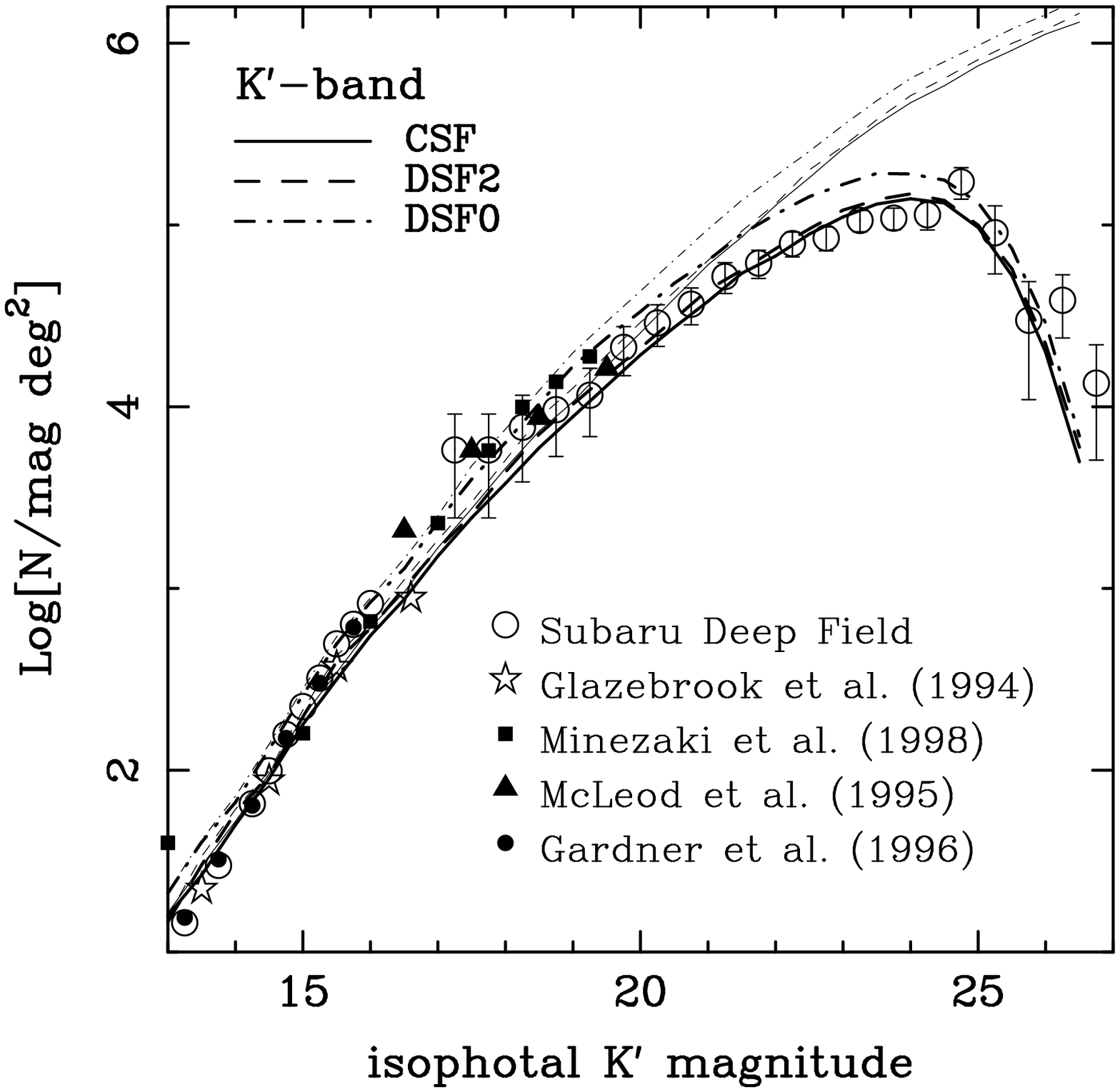}

\caption{Galaxy number counts in the $K' (2.13\mu m)$ band.  Same as
Figure \ref{fig:hdf}, but for the near-infrared counts in the SDF.  In
addition to the models of CSF (solid lines) and DSF2 (dashed lines), the
results of the DSF0 model (dot-dashed lines) are shown for comparison.
}

\label{fig:sdf}
\end{figure}

Figure \ref{fig:sdf} shows the near-infrared galaxy counts in the SDF 
($K'$).  Like the HDF counts, the models of CSF and DSF2 well reproduce 
the SDF counts.  For the purpose of comparison, the DSF0 model is also 
shown.  Because of the highest SFR at high redshift, the stellar mass 
in this model increases rapidly, making too many faint galaxies to 
reproduce the observation.  Therefore, the models of CSF and DSF2 give 
the predictions that agree with the HDF and SDF counts simultaneously.

\subsection{Redshift Distribution}
\begin{figure}
\plotone{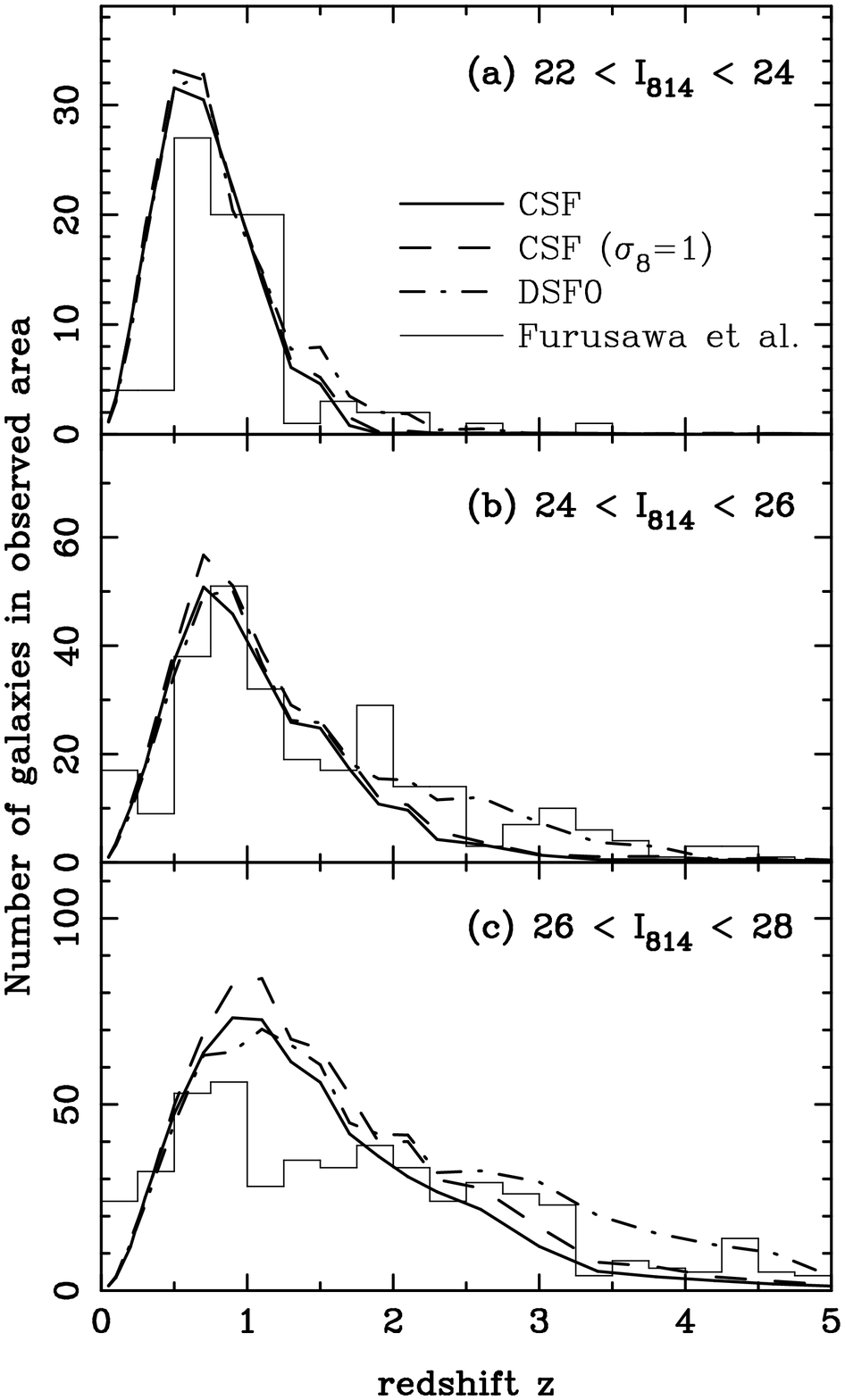}

\caption{Redshift distribution of the HDF galaxies for (a) $22\leq
I_{814}\leq 24$, (b) $24\leq I_{814}\leq 26$, and (c) $26\leq
I_{814}\leq 28$.  The thick solid, dashed and dot-dashed lines represent
the models of CSF, CSF with $\sigma_{8}=1$ and DSF0, respectively, where
the effects of dynamical response to starburst-induced gas removal are
taken into account. The histogram in each panel is the observed
photometric redshift distribution by \citet{f00}.  } 

\label{fig:hdfz}
\end{figure}

Figure \ref{fig:hdfz} shows the redshift distributions of
$I_{814}$-selected galaxies in the HDF with $22\leq I_{814}\leq 24$
({\it panel a}), $24\leq I_{814}\leq 26$ ({\it panel b}) and $26\leq
I_{814}\leq 28$ ({\it panel c}).  The number of model galaxies in each
panel is calculated over the same celestial area as the HDF.  The thick
solid, dashed and dot-dashed lines show the results of CSF, CSF with
$\sigma_{8}=1$ and DSF0, respectively.  The histogram in each panel is
the observed photometric redshift distribution by \citet{f00}, in which
they improved the redshift estimation over \citet{fly99}.  The
difference between the results of CSF and DSF0 is very small for bright
galaxies with $I_{814}\leq 26$ ({\it panels a} and {\it b}).  While the
CSF model reproduces the observed redshift distribution over an entire
range of apparent $I_{814}$-magnitudes considered, the DSF0 model
predicts too many high-redshift galaxies for $I_{814}\geq 26$ ({\it
panel c}), especially at $z\ga 3$.  This is because the shorter SF
timescale in the DSF0 model induces earlier formation of galaxies
compared with CSF.  These results are consistent with our previous
results.  We show the same model as CSF but for $\sigma_{8}=1$.  Since
the formation epoch of dark halos shifts to higher redshift, a little
more galaxies are formed at higher redshifts.  We also find that the
DSF2 model reproduces the observed redshift distribution with negligible
difference from the CSF model with $\sigma_{8}=1$.

\subsection{Isophotal Area-Magnitude Relation}\label{sec:angular}
\begin{figure}
\plotone{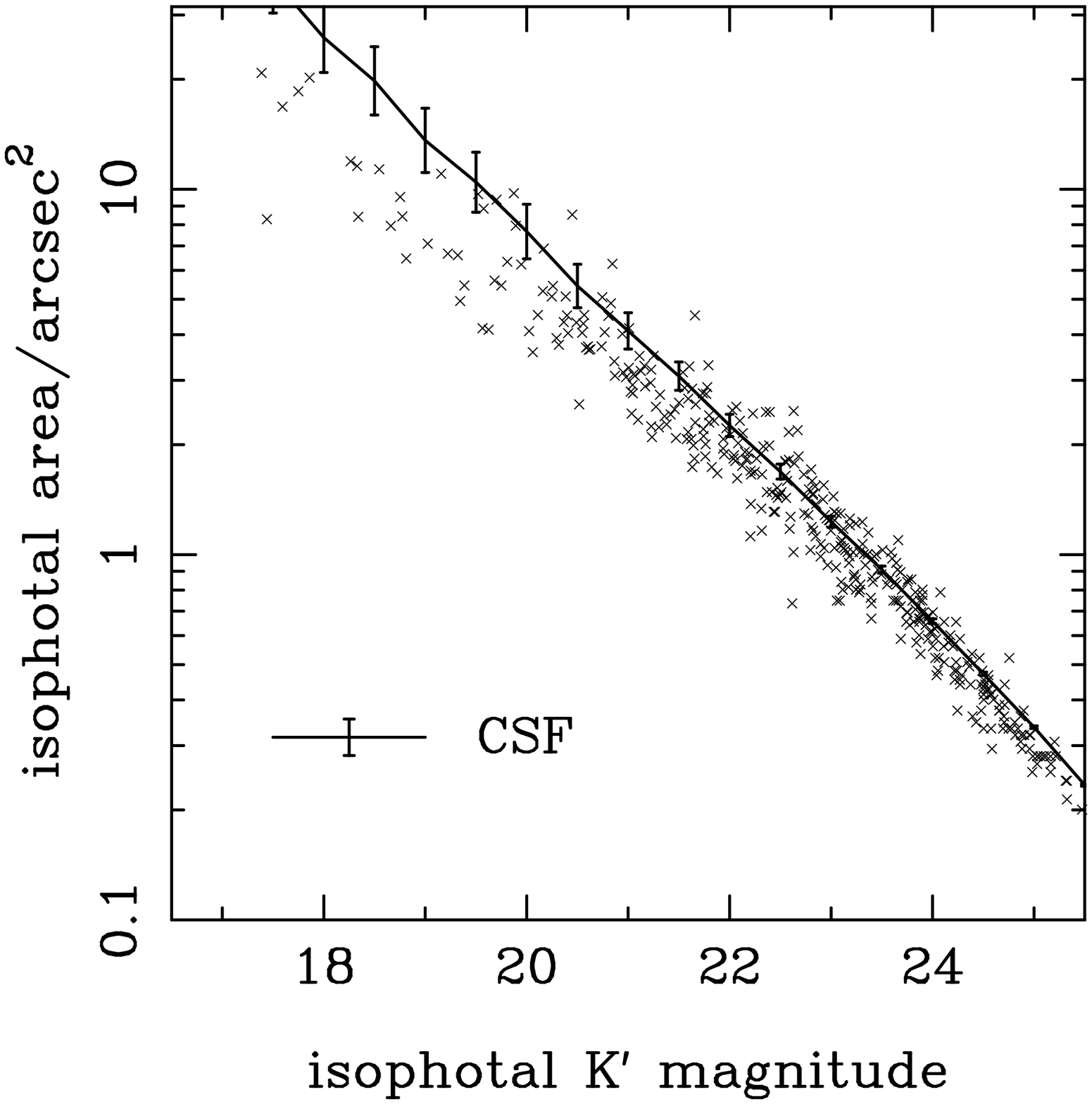}

\caption{Isophotal area of the SDF galaxies against $K'$-magnitude.  The
solid line indicates the the mean relation with errorbars of 1$\sigma$
scatter, predicted by the CSF model, where the effects of dynamical
response to starburst-induced gas removal are taken into account.  The
isophoto in this model is calculated from the observational condition
employed in the SDF survey.  The crosses indicate the observational data
from the SDF survey.  }

\label{fig:angsize}
\end{figure}

Figure \ref{fig:angsize} shows the isophotal area of galaxies plotted
against their $K'$-magnitude, for which the same observational condition
employed in the SDF survey is used to calculate the isophoto in the SF
models. The solid line indicates the the mean relation with errorbars of
1$\sigma$ scatter, predicted by the CSF model.  Results of other SF
models are almost the same, and are not shown.  The data plotted by the
crosses are those for the SDF galaxies that are detected in both the
$K'$- and $J$-bands.  As stressed in our previous papers, the selection
effects from the cosmological dimming of surface brightness of galaxies
cannot be ignored in the SAM analysis of galaxy number counts.  This
indicates that the size of high-redshift galaxies must be modeled
properly.

With the selection effects correctly taken into account, the predicted 
size should converge towards the limiting magnitude, because faint 
galaxies with larger area have surface brightnesses below the 
detection threshold and then remain undetected.  We find from this 
figure that our SAM galaxies well reproduce the observed area-magnitude 
relation, and are consistent with the SDF galaxies, only when the 
selection bias against faint galaxies with high redshift and/or low 
surface brightness is taken into account in the analysis.

\subsection{Tully-Fisher Relation (TFR)}
\begin{figure}
\plotone{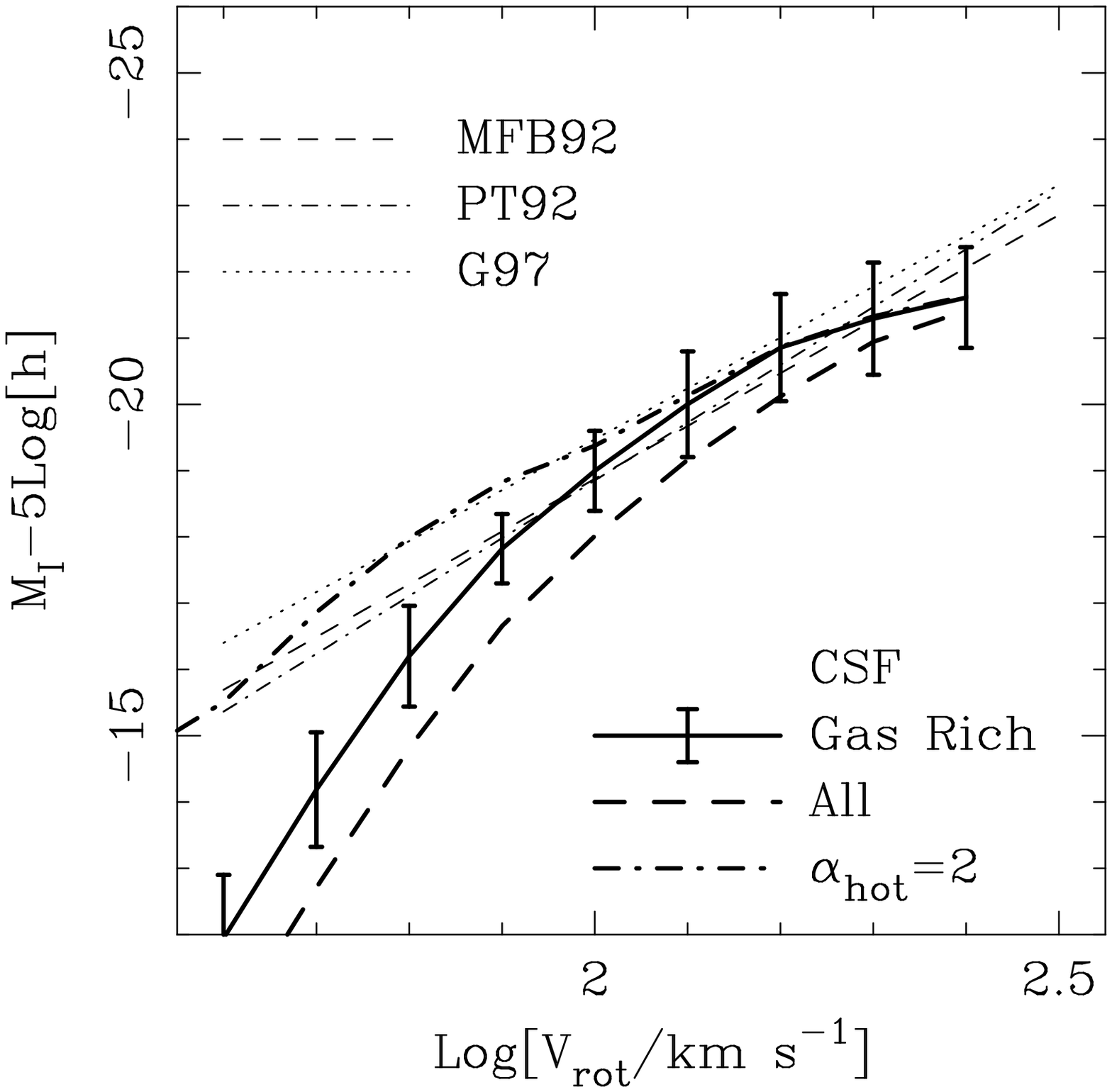}

\caption{$I$-band Tully-Fisher relation of spiral galaxies.  Variants of
the CSF model with the effects of dynamical response are shown for the
cases of only gas rich spirals having the cold gas more than 10\% in
mass (thick solid line), all spirals with our standard set of parameters
(thick dashed line), and gas rich spirals with a smaller value of
$\alpha_{\rm hot}=2$ (thick dot-dashed line).  Errorbars indicate
1$\sigma$ scatter around the predicted mean.  The observed TFRs are
shown as a mean relation given by \citet{mfb92} (thin dashed line),
\citet{pt92} (thin dot-dashed line), and \citet{g97} (thin dotted line).
}

\label{fig:tf}
\end{figure}

Figure \ref{fig:tf} shows the $I$-band TFR.  The thick solid line 
indicates the predicted TFR with errorbars of 1$\sigma$ scatter
for gas-rich spirals in the CSF model, having more than 10\% 
mass fraction of cold gas in the total galactic mass, that is, 
$M_{\rm cold}/(M_{*}+M_{\rm cold})\geq 0.1$.  This criterion is 
the same as that used by \citet{c00}.  The thick dashed line is 
for all spirals.  The tendency that gas-rich spirals for a given 
magnitude are always brighter than gas-poor spirals has already 
been shown in \citet{c00}.  Results of other SF models are not 
shown because of negligible difference. 

The thin dashed, dot-dashed and dotted lines are the best-fit 
results to the observed TFRs by \citet{mfb92}, \citet{pt92}, and 
\citet{g97}, respectively.  We assume that the line-width $W$ is 
simply twice the disk rotation velocity $V_{\rm rot}$ as usual.  
For model galaxies, we set $V_{d}$ to be equal to $V_{\rm rot}$, 
as mentioned in \S\S\ref{sec:response}.

The predicted TFR slope for galaxies with small rotation velocity
$V_{\rm rot}\la 80$km is steeper than that observed, while the predicted
slope and magnitude for galaxies with larger velocity agree well with
the observations.  The TFR slope is determined by $\alpha_{\rm hot}$,
because this parameter relates the mass fraction of stars in each
galaxy, which gives luminosity, to the rotation velocity.  We use
$\alpha_{\rm hot}=4$ throughout as a standard, but smaller $\alpha_{\rm
hot}$ gives shallower slope. For the purpose of comparison, we also plot
the CSF model with $\alpha_{\rm hot}=2$ by the dot-dashed line.  In this
case the slope becomes closer to that observed.  Although \citet{c00}
adopted $\alpha_{\rm hot}=2$ and claimed agreement of their model with
the observed TFRs, we found that this model predicts too many
high-redshift galaxies to be reconciled with the observed number counts.
Moreover, mean metallicity of stars in dwarf spheroidals is too high to
be consistent with their observed metallicities, as shown in
\S\S\ref{sec:metallicity}.  In order to reproduce both observations, it
might be worth relaxing our assumptions in estimating the disk rotation
velocity and/or other physical processes such as SN feedback.  For
example, the SN feedback parameter $\beta$, which is assumed to be
constant, might depend on whether star formation is either continuous or
burst-like.  Furthermore, it might evolve with redshift.  Dynamical
response to gas removal, of course, provides a promising effect on disk
rotation velocity.  These possibilities should therefore be investigated
in more detail.

\subsection{Color-Magnitude Relation (CMR)}\label{sec:cmr}
\begin{figure}
\plotone{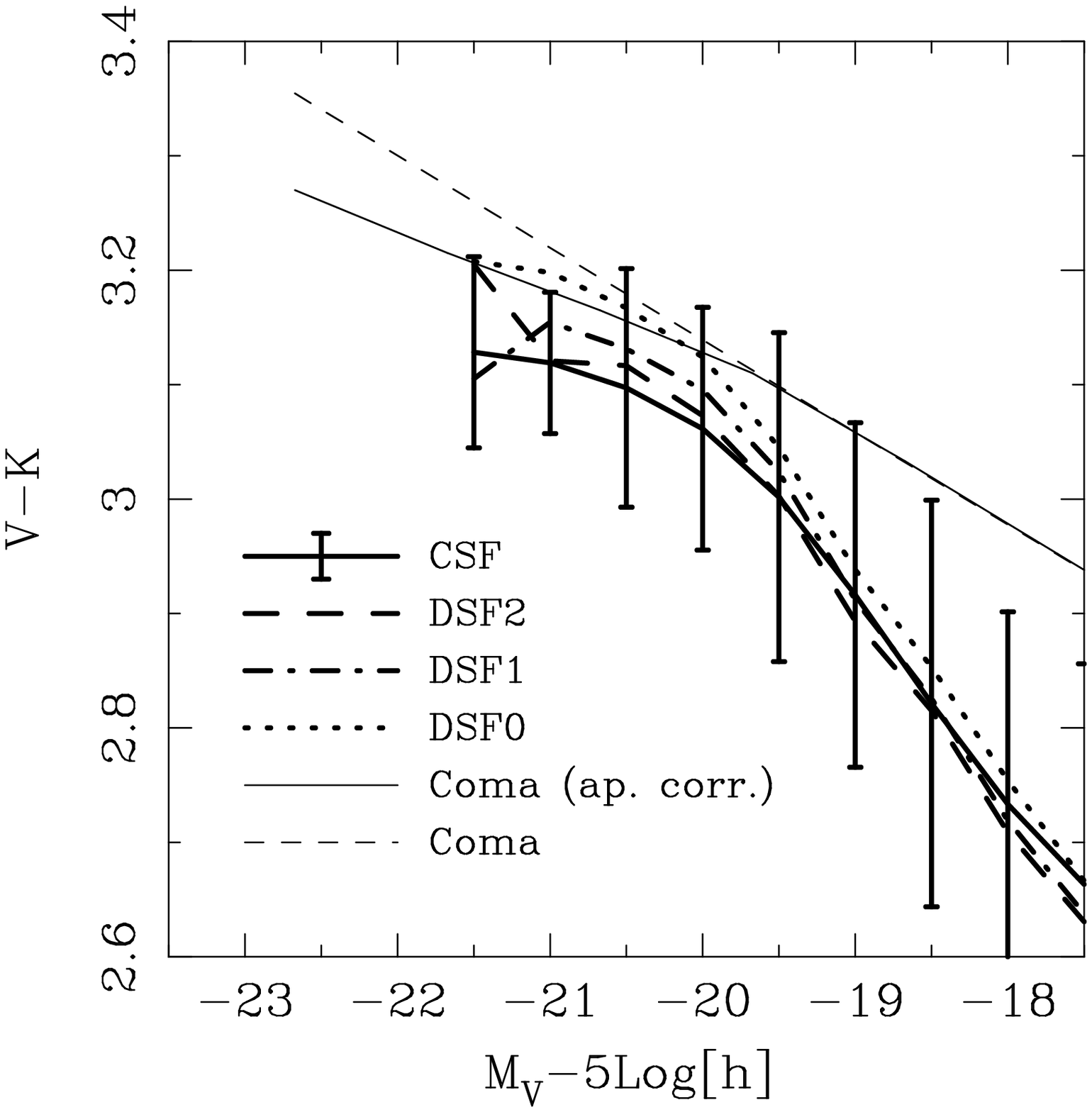}

\caption{$V-K$ color-magnitude relation of elliptical galaxies in 
clusters.  Such galaxies are sampled from the clusters having
$V_{c}=10^{3}$ km s$^{-1}$, and the theoretical CMRs are shown
for the models of CSF (thick solid line), DSF2 (thick dashed line), 
DSF1 (thick dot-dashed line), and DSF0 (thick dotted line), where 
the effects of dynamical response to starburst-induced gas removal 
are taken into account.  The observed CMR for the Coma cluster is
shown as a mean relation by \citet{ble92} (thin dashed line), and
the aperture-corrected CMR for the same cluster by \citet{kaba98}
(thin solid line).
}
\label{fig:cmr}
\end{figure}

Figure \ref{fig:cmr} shows the $V-K$ color versus magnitude relation 
of cluster elliptical galaxies embedded in extended dark halo 
with $V_{\rm circ}=10^{3}$km s$^{-1}$.  The thick solid, dashed, 
dot-dashed, and dotted lines represent the predicted CMRs in the SF 
models of CSF, DSF2, DSF1, and DSF0, respectively.  The CMR for each 
model is obtained by averaging 50 realizations.  Errorbars to the
CMR denote the $1\sigma$ uncertainties in the DSF2 model, which are 
comparable to CSF and DSF1 but is slightly larger than DSF0.  The thin 
dashed line is the observed CMR for galaxies in the Coma cluster by 
\citet{ble92} and the thin solid line is the same but for the 
aperture-corrected CMR by \citet{kaba98}.

All the SF models are not reconciled with the observed CMR, although 
the colors at $M_{V}-5\log h\simeq -20$ are consistent with the 
observation.  This is a consequence of adopting the SN feedback with 
$(\alpha_{\rm hot}, V_{\rm hot})=(4, 180{\rm km~s}^{-1})$, together
with the satellite-satellite merger.  As discussed by \citet{kc98} 
and \citet{ng01}, a combination of smaller $\alpha_{\rm hot}
(\simeq 2)$ and larger $V_{\rm hot}$ is known to give a better fit 
to the observed CMR.  

\begin{figure}
\plotone{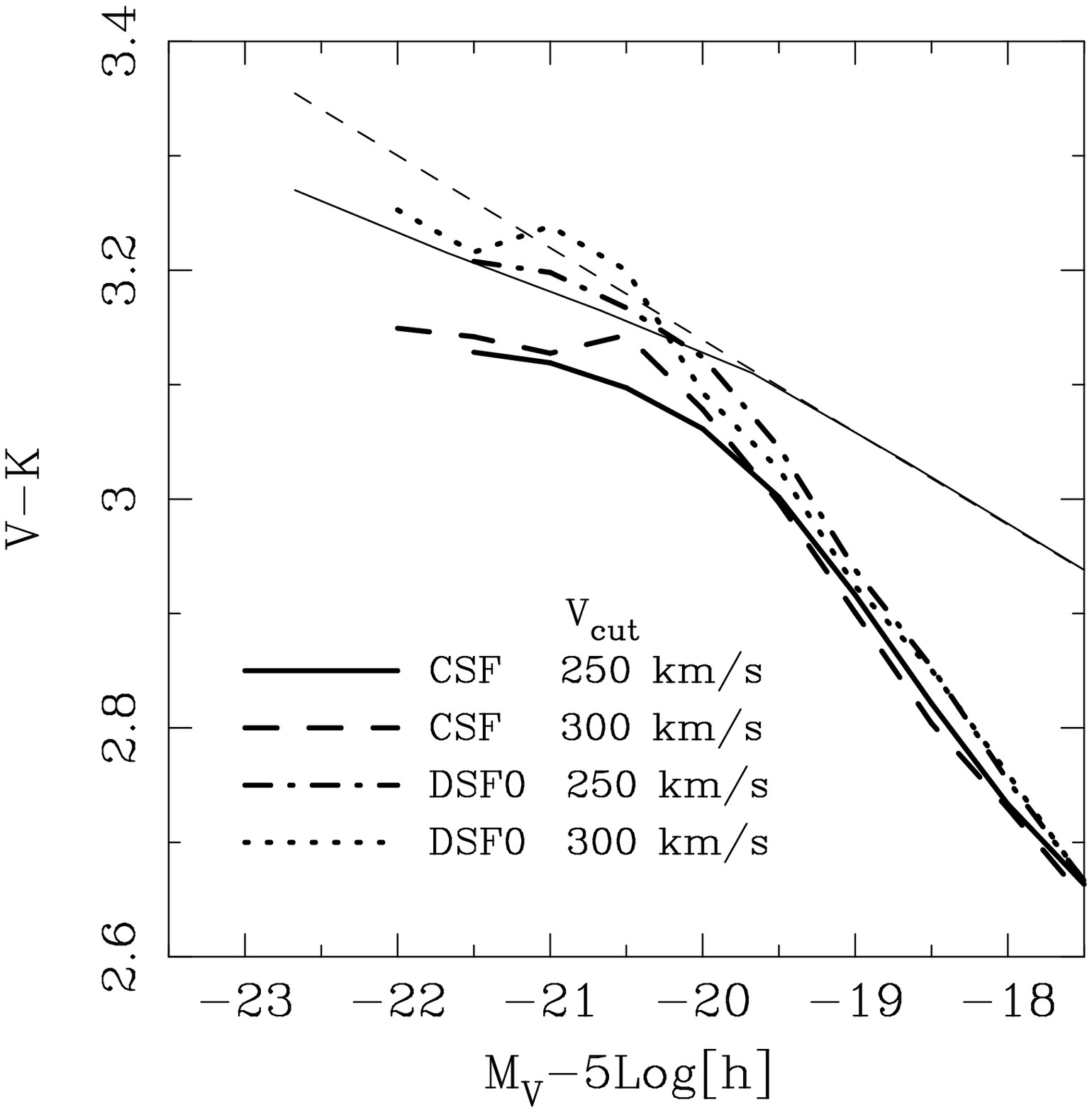}

\caption{$V-K$ color-magnitude relation of elliptical galaxies in
clusters. Same as Figure \ref{fig:cmr}, but for the effect of changing
$V_{\rm cut}$.  The thick solid and dashed lines show the theoretical
CMRs in the CSF model with $V_{\rm cut}=250$ (our standard choice) and
300 km s$^{-1}$, respectively.  The dot-dashed and dotted lines
similarly show these two cases for the DSF0 model, respectively.  }
\label{fig:cmr2}
\end{figure}

Like the TFR, the faint-end slope of the CMR is determined by 
$\alpha_{\rm hot}$, because the CMR is primarily a metallicity 
sequence \citep{ka97} and chemical enrichment of dwarf galaxies 
is determined by $\alpha_{\rm hot}$.  The bright-end slope of 
the CMR is nearly flat, because the SN feedback becomes negligible.
If we adopt a large value of $V_{\rm hot}\simeq 280$ km s$^{-1}$, 
this flat region moves brightwards.  Thereby, an expected slope 
from $\alpha_{\rm hot}$ is realized over an almost entire range of 
magnitudes.  Additionally, the satellite-satellite merger makes dwarf 
spheroidal galaxies brighter, while keeping their color unchanged.  
This shifts the CMR bluewards when seen at faint magnitudes.

With these considerations, we can improve the fit to the observed CMR
by adjusting the SN feedback-related parameters ($\alpha_{\rm hot}$, 
$V_{\rm hot}$), and the merger-related parameter $f_{\rm mrg}$. 
However, appropriate choice of their values that could explain the 
observed CMR seems to invalidate other successes of our SAM.  This 
difficulty is clearly the area of future investigation. 

Furthermore, we find that a bright portion of the CMR is also affected
by the cooling cutoff $V_{\rm cut}$.  Figure \ref{fig:cmr2} shows the
$V_{\rm cut}$-dependence of CMR.  The solid and dashed lines represent
the CSF model with $V_{\rm cut}=250$ and 300 km s$^{-1}$, respectively.
The dot-dashed and dotted lines are the same but for the DSF0 model.  If
we adopt larger $V_{\rm cut}$, the metallicity becomes larger because
chemical enrichment continues until lower redshift. Note that the
difference between the models of CSF and DSF0 is caused mainly by the
difference of mean stellar age, that is, stars in the DSF0 model were
born earlier and thus older than CSF.  Since the CMR is sensitive to
$V_{\rm cut}$, more knowledge of the gas cooling is needed to fix the
cutoff on cluster scales.

This situation might be improved by taking the following procedures.  
First is to use a lower value of $\sigma_{8}$ below unity as recent 
observations suggest $\sigma_{8}\simeq 0.8$ \citep[e.g.,][]{s03}. 
This value provides a statistically late epoch for density fluctuations 
to collapse.  Therefore, a large value of $V_{\rm cut}$ does not result
in the formation of monster galaxies. Second is the aperture effects. 
\citet{kg03} showed the importance of aperture effects by using their 
chemo-dynamical simulations.  The predicted $V-K$ colors of their giant 
elliptical galaxies are on the average 3.2 at $M_{V}-5\log h\simeq -22$ 
and agree well with the observed colors within 5 kpc aperture. However, 
the observed $V-K$ color within 99kpc aperture is nearly equal to 3.0, 
which is even bluer than our predicted color at the same magnitude.  
Third is the effects of UV background.  \citet{ng01} showed that the 
photoionization by the UV background has a similar effect to the SN 
feedback.  Therefore, introduction of the photoionization is equivalent 
to adopting larger $V_{\rm hot}$.  Further investigation on the CMR as 
well as the TFR are needed along these lines.

It should be noted that \citet{on03} have successfully reproduced the CMR
slope by using a SAM combined with an $N$-body simulation.  Their SF model 
corresponds to CSF with $\alpha_{\rm hot}=2$ and $V_{\rm hot}=200$ km s$^{-1}$.
As explained above, however, adopting this small value of $\alpha_{\rm hot}$ 
fails to explain other observations unless the merger strength is adjusted. 
In this sense, high resolution $N$-body simulations is highly awaited to 
follow a full trace of merging histories of progenitor halos of clusters.

\section{Cosmic Star Formation History}\label{sec:Madau}
\begin{figure}
\plotone{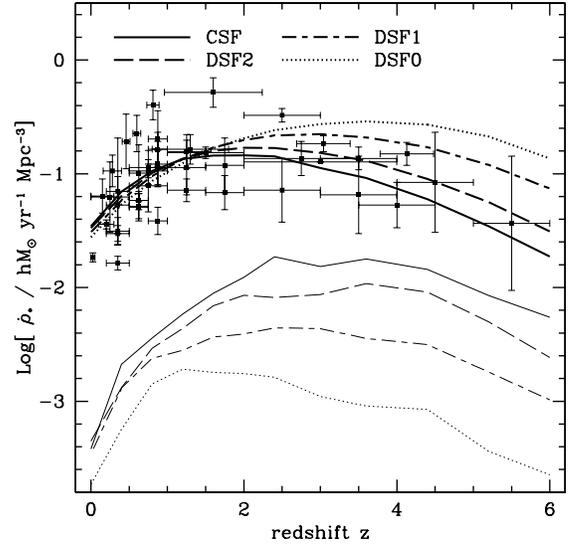}

\caption{Cosmic Star Formation Histories.  The solid, dashed, dot-dashed
 and dotted lines denote CSF, DSF2, DSF1 and DSF0, respectively.  The
 thick lines indicate the total SFR and the thin lines the SFR only for
 starburst.  Symbols with errorbars are observational data compiled by
 \citet{aygm02}.  }

\label{fig:Madau}
\end{figure}

Recently the cosmic SF history, which is a plot of SFR in a comoving
volume against redshift, is widely used to examine the global SF history
\citep{m96} and also in SAM analyses \citep{bcfl98, spf01}.  Since the
SFR is very sensitive to the SF timescale, in Figure \ref{fig:Madau} we
plot the redshift evolution of cosmic SF rate for the four models with
different SF timescales.  Thick lines denote the total SFR and thin
lines the SFR only for starburst.  Symbols with errorbars indicate
observational SFRs compiled by \citet{aygm02}.  While there are large
scatters between individual data points, the CSF and DSF2 models broadly 
agree with the observations.

As expected from the redshift dependence of SF timescale, the CSF model
predicts a maximum SFR at relatively low redshift, $z\la 2$, and the
redshift at which this maximum occurs becomes larger, in order from DSF2
(dashed line) via DSF1 (dot-dashed line) to DSF0 (dotted line) as the SF
timescale becomes shorter at high redshift (see Figure \ref{fig:sfr}).

The fraction of cold gas available for starburst becomes larger for
longer SF timescale at high redshift.  Because of the same reason as
above, the CSF model by the thin solid line gives the largest fraction
of cold gas at major merger and therefore the highest SFR for starburst
among the four models under consideration.  Accordingly, the dynamical
response is the most effective in the CSF model.

\section{Cooling Diagram Revisited}
\citet{s77} and \citet{ro77} proposed the so-called cooling diagram, that 
is, the distribution of galaxies on the density versus temperature diagram.
Since then, a number of authors used this diagram as crucial constraints
on the formation of galaxies in the framework of monolithic cloud collapse 
scenario \citep[e.g.,][]{f82, bfpr84}.  For example, characteristic mass 
of galaxies can be evaluated, because the evolutionary path of gas clouds in 
this diagram gives an initial condition of density and temperature expected
from density fluctuation spectrum in the early universe.  

\begin{figure}
\plotone{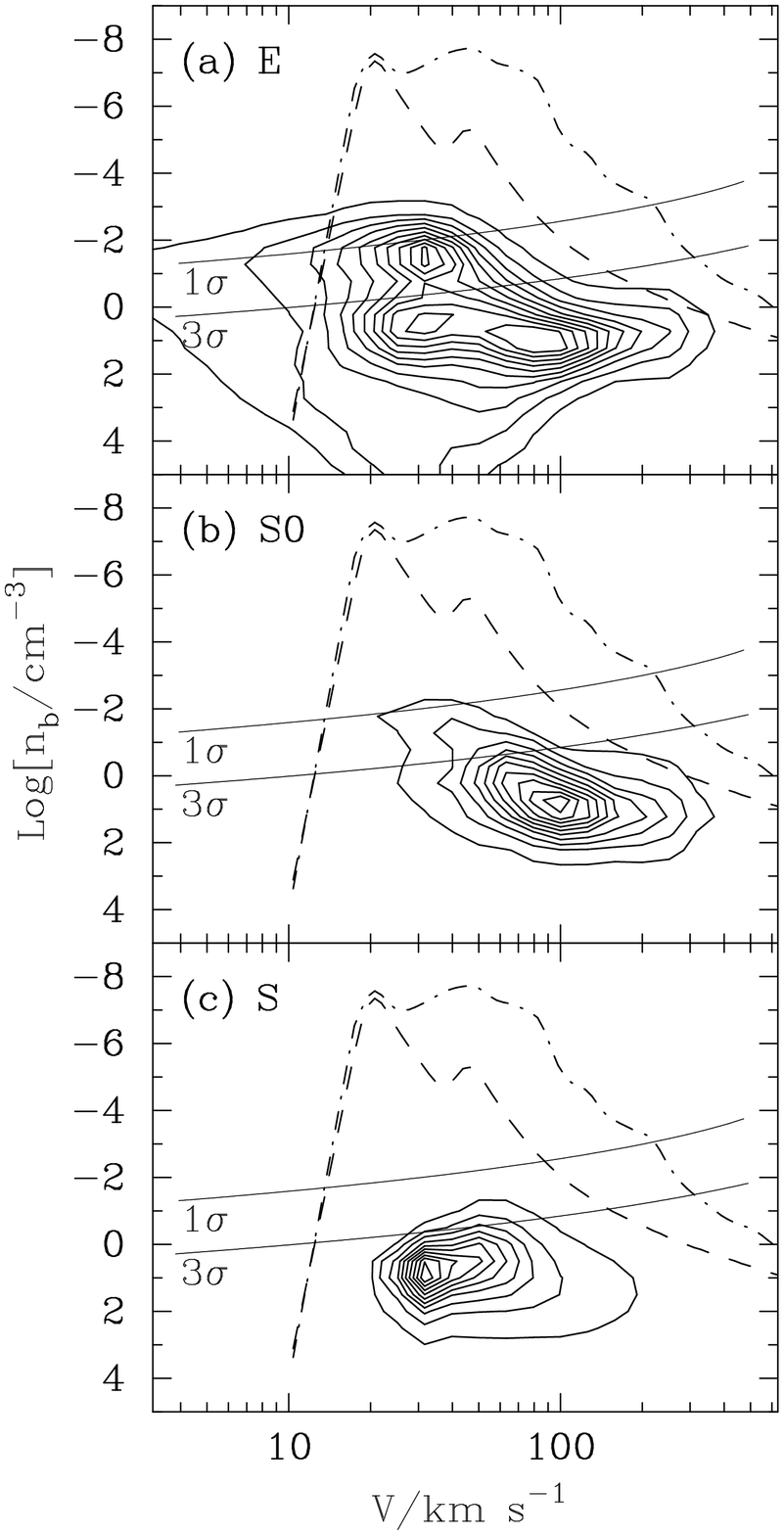}

\caption{Cooling diagram for (a) elliptical galaxies, (b) lenticular
galaxies, and (c) spiral galaxies.  Contours indicate the theoretical
distribution predicted by the CSF model with surface brightness cutoff
at $\mu_{e,B}=26.5$ and the effects of dynamical response taken into
account in the analysis.  The levels of contours are the same as in
Figure \ref{fig:rad1}.  Top panel shows a broad distribution of
elliptical galaxies consisting of dwarf ellipticals on the left and
normal/giant ellipticals on the right, which are bifurcated reflecting a
variety of star formation and merger histories.  The dashed line
represents $\tau_{\rm cool}=\tau_{\rm grav}$ for the gas of primordial
chemical composition, and the dot-dashed line for the gas of solar
chemical composition.  The thin lines represent two sequences of
collapse of overdense regions with 1$\sigma$ and 3$\sigma$ fluctuations
against CDM spectrum.  } \label{fig:cd1}
\end{figure}

Figure \ref{fig:cd1} shows the contours of galaxy distribution in the 
cooling diagram for ellipticals ({\it top panel}), lenticulars ({\it 
middle panel}), and spirals ({\it bottom panel}).  The velocity on the 
horizontal axis indicates the velocity dispersion of bulge component for 
elliptical and lenticular galaxies, and the disk rotation velocity for 
spiral galaxies.  The baryon density on the vertical axis is estimated 
by dividing the baryonic mass in individual galaxies by their volume.  
We simply assume a sphere of effective radius for elliptical and lenticular
galaxies, and a cylinder of scale height being one tenth of the effective 
radius for spiral galaxies. The contours are plotted excluding low surface 
brightness galaxies with $\mu_{e,B}\geq 26.5$.

\begin{figure}
\plotone{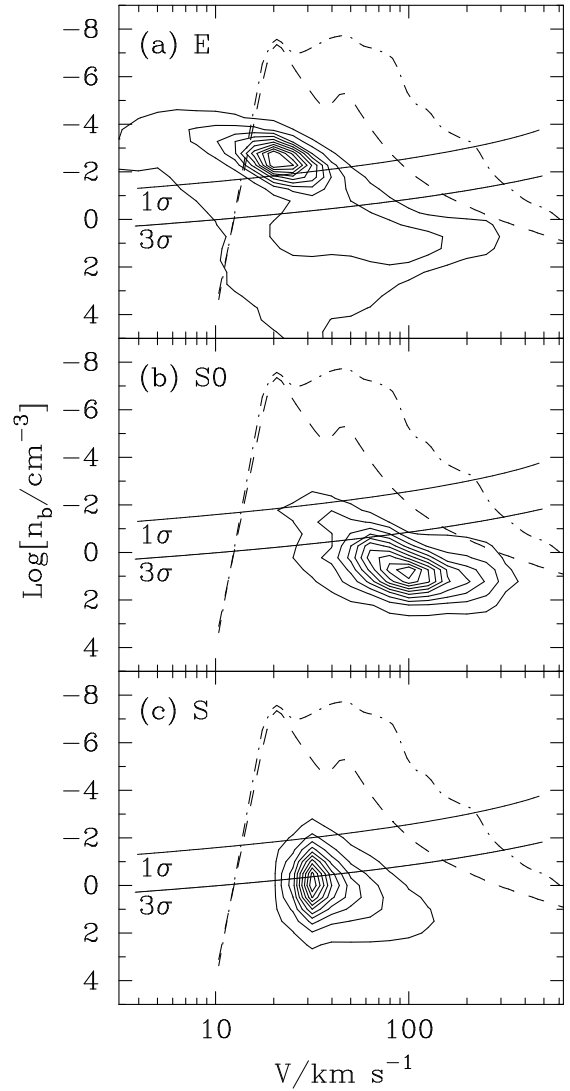}
\caption{Cooling diagram.  Same as Figure \ref{fig:cd1}, but for all 
galaxies without imposing any selection bias against low surface 
brightness.
}
\label{fig:cd2}
\end{figure}

The dashed and dot-dashed lines represent the cooling curves that the 
cooling timescale is equal to the gravitational free-fall timescale 
$\tau_{\rm cool}=\tau_{\rm grav}$ for the cases of primordial and solar 
compositions, respectively.  These timescales are
\begin{eqnarray}
\tau_{\rm cool}&=&\frac{3}{2}\frac{\rho_{\rm gas}}{\mu m_{\rm p}}\frac{kT}{n_{\rm e}^{2}\Lambda(T,Z_{\rm gas})},\\
\tau_{\rm grav}&=&\left(24\pi G\rho_{\rm tot}\right)^{-1/2},
\end{eqnarray}
where $n_{\rm e}$ is the electron number density, $\rho_{\rm gas}$ is
the gas density, $\rho_{\rm tot}$ is the total mass density including
dark matter, and $\Lambda(T,Z_{\rm gas})$ is the cooling function
depending on both the gas temperature $T$ and metallicity $Z_{\rm gas}$.
Wherever necessary below, we transform the gas temperature into velocity
dispersion based on the virial theorem, and identify the gas density
with baryon density. The thin solid curves represent the sequences of
galaxies originated from the CDM density fluctuations with density
contrast of $1\sigma$ and 3$\sigma$, respectively.  It is evident from
this figure that our SAM galaxies are distributed within a region of
$\tau_{\rm cool}<\tau_{\rm grav}$ and their morphologies are distinctly
segregated from each other, as observed in this diagram.

Figure \ref{fig:cd2} shows the distribution for all galaxies without 
imposing any selection bias against low surface brightness.  
Comparison of Figures \ref{fig:cd1} and \ref{fig:cd1} indicates that 
many dwarf galaxies have low surface brightness and are distributed
towards a region of $\tau_{\rm cool}>\tau_{\rm grav}$ characterized by 
low baryon density and low circular velocity.  This extended distribution 
is prominent only for elliptical galaxies for which the starburst-induced 
gas removal followed by dynamical response has the maximal effect.  Such 
galaxies of low surface brightness would have been detected in recent 
ultradeep surveys where the detection threshold is set below $\mu_{e,B}=26.5$.

\section{Summary and Conclusion}
We have investigated the formation and evolution of galaxies in the
context of the hierarchical clustering scenario by using the Mitaka
model, or our SAM in which the effects of dynamical response on size and
velocity dispersion of galaxies are explicitly taken into account,
according to the formula by \citet{ny03} for galaxies consisting of
baryon and dark matter.  This paper is therefore an extension of
previous analyses in the context of monolithic cloud collapse scenario
\citep{ds86, ya87}.

A $\Lambda$-dominated flat universe, which is recently recognized as a 
standard, is exclusively used here.   The investigation mainly focuses 
on elliptical galaxies which are assumed to be formed by major merger
and starburst-induced gas removal followed by dynamical response of 
the systems.  While a mass fraction of removed cold gas, after heated, is 
determined by the circular velocity of galaxies similar to the traditional 
collapse models, the total amount of cold gas depends on the formation 
history of galaxies which is realized by the Monte Carlo method based 
on the power spectrum of density fluctuation predicted by the CDM model.  

The Mitaka model, which we have constructed in this paper, is found to
reproduce a wide variety of observed characteristics of galaxies, 
particularly their scaling relations among various observables such as 
magnitude, surface brightness, size, velocity dispersion, mass-to-light 
ratio and metallicity.  This strongly supports the CDM cosmology and 
merger hypothesis of elliptical galaxy formation even on scales of dwarf 
galaxies.

Most of model parameters related to star formation, SN feedback and
galaxy merger are constrained by the local luminosity function and mass
fraction of cold gas in spiral galaxies.  As an extra parameter to be
furthermore constrained, we have examined redshift dependence of SF
timescale through comparison among the four SF models denoted by CSF,
DSF0, DSF1, and DSF2.  The CSF refers to a constant SF timescale against
redshift and has been suggested to be consistent with observations of
galaxy number counts \citep{ntgy01, nytg02}, quasar luminosity function
\citep{kh00, eng03}, and number and metallicity evolution of damped
Ly-$\alpha$ systems \citep{spf01, ongy02}.  The DSF0 refers to a SF
timescale simply proportional to dynamical timescale. The DSF1 and DSF2
are intermediate between the CSF and DSF0 (see Figure \ref{fig:sfr}).
It is found that the DSF0 model fails to reproduce observed properties
of local dwarf spheroidals and galaxy counts.  Among the above four SF
models, the CSF model has the longest SF timescale at high redshift and
therefore the largest amount of cold gas at that epoch.  Accordingly,
the fraction of removed gas during starburst is the largest in the CSF
model, giving many extended galaxies in agreement with observation.  In
this sense, the model of constant star formation (CFS) or at most mildly
evolving star formation (DSF2) is favorable.  Note that {\it dominant
dark halo}, in which size and velocity dispersion do not change during
gas removal as considered by \citet{ds86}, apparently explains the
observed scatter of size.  However, arguments from physical ground
indicate that the dynamical response should play a significant role in
forming galaxies with low velocity dispersion of $\sigma_{0}(\rm 1D)\la
10$km s$^{-1}$, as actually observed in the Local Group.

There are some areas of further improvements in the current Mitaka 
model.  Because of strong dependence of SN feedback on circular velocity
assuming $\alpha_{\rm hot}=4$, dwarf galaxies in our SF models are too 
faint to agree with expected magnitude from observed TFRs.  Furthermore, 
because of the same reason, dwarf galaxies are too blue to agree with 
expected colors from observed CMRs for cluster elliptical galaxies.  
On the other hand, such large value of $\alpha_{\rm hot}$ stops chemical 
and photometric evolution of elliptical galaxies at early epochs and is 
required to explain their observed low stellar metallicity as well as 
the galaxy number counts in the HDF and SDF.  In order to settle these 
contradictions, some new ingredients need to be introduced in the SAM 
analysis.  At least, for example, we must know how efficient the gas 
cooling is in massive dark halos and whether SN feedback at starburst 
works in the same way as that in disks. 
 
We adopted a fitting mass function of dark halos by \citet{yny03} for
given redshift instead of the often used PS mass function.  This YNY
mass function is slightly different from those by \citet{st99} and
\citet{j01}, but is confirmed to provide a better fit to recent $N$-body
results given by \citet{yy01} and \citet{y02}, which is discussed in
more detail in a separate paper \citep{yny03}.  Since the number density
of dark halos affects SN feedback-related parameters to be chosen, it is
very important to establish the mass function of dark halos,
particularly on small mass scales, by high resolution $N$-body
simulation.  Although such SAM analyses have just begun
\citep[e.g.,][]{bpfbj01, h03a, h03b}, further investigation will
obviously be required.  We are upgrading the Mitaka model accommodated
with full high resolution $N$-body simulations, and the results will be
given elsewhere in the near future.

\begin{figure}
\plotone{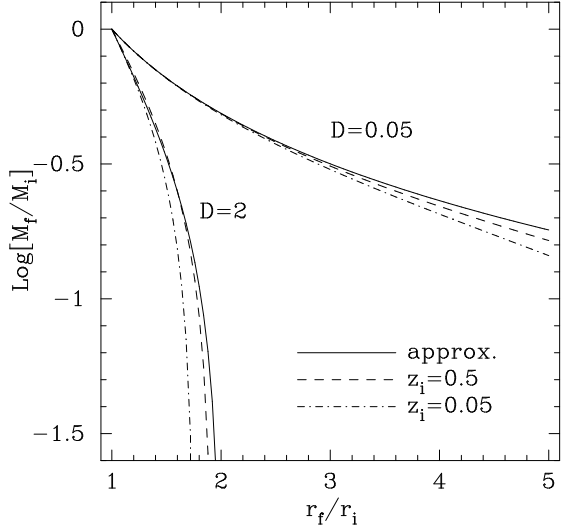}
\caption{Dynamical response of galaxy size to removed mass from 
the system.  The approximate results given by equation (A9) 
are shown for $D=0.05$ and 0.2 (solid lines).  For each value of
$D$, the exact results given by equation (\ref{eqn:full}) are 
shown for $z_{i}=0.5$ (dashed lines) and 0.05 (dot-dashed lines).
}  
\label{fig:approx}
\end{figure}

\acknowledgments

We thank Takashi Okamoto and Naoteru Gouda for useful suggestions.  
The authors are grateful to the anonymous referee for helpful comments 
to improve this paper.  This work has been supported in part by the 
Grant-in-Aid for the Center-of-Excellence research (07CE2002) of the 
Ministry of Education, Science, Sports and Culture of Japan.  MN 
acknowledges support from a PPARC rolling grant for extragalactic 
astronomy and cosmology.

\appendix
\section{Dynamical Response to Gas Removal}
We show the dynamical response of size and velocity dispersion 
to supernova-induced gas removal in the two-component galaxies 
consisting of baryon and dark matter.  The details are given in 
\citet{ny03}.

All elliptical galaxies in this paper have a density profile similar
to the Jaffe model \citep{jaffe},
\begin{equation}
 \rho(r)=\frac{4\rho_{b}r_{b}^{4}}{r^{2}(r+r_{b})^{2}},
\end{equation}
where $\rho_{b}$ and $r_{b}$ are the characteristic density and 
radius, respectively.  This profile well approximate the de Vaucouleurs 
$r^{1/4}$ profile for stars.  The effective (half-light) radius 
$r_{e}$ defined in the projected surface is related to the half-mass 
radius $r_{b}$ in three dimensional space as $r_{e}=0.744r_{b}$ 
\citep{ny03}.  For the distribution of dark matter, the singular 
isothermal distribution ($\rho_{d}\propto r^{-2}$) is considered.  
We found that the Navarro-Frenk-White profile \citep{nfw97} provides 
almost the same results as those given by singular isothermal sphere.  
Note that according to recent high resolution $N$-body simulations, 
which have reveals that many substructures survive even in virialized 
halos, we assume that subhalos exist as underlying gravitational 
potential for satellite galaxies.  As mentioned in \S\S\ref{sec:tidal}, 
these subhalos are tidally truncated.

Because the derivation is long and complicated, we describe it only
briefly.  Just after the merger, the system is virialized immediately.  
Then the starburst occurs and a part of gas is gradually removed.  
The size and velocity dispersion change following the gas removal 
adiabatically. During the gas removal, the dark matter distribution 
is assumed to be not affected.

In the followings we derive useful simplified formulae for the dynamical
response.  As shown in Appendix B of \citet{ny03}, the relationship
between density and size for the adiabatic gas removal is given by
\begin{equation}
 Y=\frac{h(y_{i},z_{i})p(z_{f})+q(z_{f})}{y_{i}},\label{eqn:response}
\end{equation}
where $Y\equiv y_{f}/y_{i}$, $y_{i}=\rho_{i}/\rho_{d}$,
$z_{i,f}=r_{i,f}/r_{d}$, and the function $h, p$ and $q$ are
\begin{eqnarray}
h(y,z)&=&yz^{4}+\frac{1}{2}[-z+(1+z)\ln (1+z)],\\
p(z)&=&\frac{1}{z^{4}},\\
q(z)&=&-\frac{1}{2z^{4}}[-z+(1+z)\ln (1+z)].
\end{eqnarray}
In order to obtain the size after gas removal, we need to know the
inverted function of the above and it will be very complicated.
Therefore, we approximate the above function.  The equation can be
reduced to the following expression,
\begin{equation}
 Y=\frac{1}{R^{4}}+\frac{U_{\rm iso}(R,z_{i})}{2y_{i}z_{i}^{4}R^{4}},\label{eqn:full}
\end{equation}
where $R=z_{f}/z_{i}$.
Expanding $U_{\rm iso}(R,z_{i})$ around $R=1$ and $z_{i}=0$ and picking out the
lowest order term, we obtain
\begin{equation}
 Y=\frac{1}{R^{4}}-\frac{D}{2}\frac{R-1}{R^{4}},
\end{equation}
where $D=1/y_{i}z_{i}^{2}$.  Defining the ratio of final to initial masses, 
${\cal M}=M_{f}/M_{i}=YR^{3}$, the above
equation is transformed to
\begin{equation}
 {\cal M}=\frac{1}{R}-\frac{D}{2}\frac{R-1}{R}.
\end{equation}
Then, inverting this, we obtain
\begin{equation}
 R=\frac{1+D/2}{{\cal M}+D/2}.
\end{equation}
By using the virial theorem, the velocity dispersion is
\begin{equation}
 \frac{\sigma_{f}}{\sigma_{i}}=\sqrt{\frac{YR^{2}+Df(z_{f})/2}{1+Df(z_{i})/2}},
\end{equation}
where
\begin{equation}
 f(z)=\frac{\ln(1+z)}{z}+\ln\left(1+\frac{1}{z}\right).
\end{equation}

Figure \ref{fig:approx} shows the relation between ${\cal M}$ and
$R$ based on this approximate formula with $D=0.05$ and 2 (solid 
lines).  For each value of $D$, the exact solution is also shown 
for two cases of $z_{i}=0.05$ (dot-dashed line) and 0.5 (dashed line). 
Clearly our approximate relation follows the exact solution quite 
well.

\end{document}